\documentclass[aps,twocolumn,showpacs,preprintnumbers,nofootinbib,prd,superscriptaddress,groupedaddress,10pt]{revtex4-1}

\makeatletter
\def\l@subsubsection#1#2{}
\def\l@subsubsubsection#1#2{}
\makeatother

\usepackage{subfigure}
\usepackage{graphicx,amssymb,amsmath,amsthm,amsfonts,epsfig,epsf}
\usepackage[usenames]{color}
\usepackage{epstopdf}
\definecolor{darkred}{rgb}{0.5,0,0}

\usepackage{array}
\usepackage{afterpage}
\usepackage{bm}
\usepackage{dcolumn}
\usepackage[utf8]{inputenc}
\usepackage{latexsym}
\usepackage{rotating}
\usepackage{longtable}

\usepackage{enumerate}
\usepackage{tensor,multirow}
\usepackage{url}
\usepackage{placeins}
\usepackage[linktocpage]{hyperref}
\usepackage{float}
\usepackage{graphicx}

\newcounter{mnotecount}[section]

\renewcommand{\themnotecount}{\thesection.\arabic{mnotecount}}

\newcommand{\mnote}[1]%{}%
{\protect{\stepcounter{mnotecount}}$^{\mbox{\footnotesize
$%\!\!\!\!\!\!\,
\bullet$\themnotecount}}$ \marginpar{%\color{red}%
\raggedright\tiny\em
$\!\!\!\!\!\!\,\bullet$\themnotecount: #1} }

\begin{document}

\title{Superradiant instability of charged scalar fields in higher-dimensional Reissner-Nordstr\"om-de Sitter black holes}

\author{
Kyriakos Destounis
%Rodrigo Vicente$^{1}$
}
\affiliation{CENTRA, Departamento de F\'{\i}sica, Instituto Superior T\'ecnico -- IST, Universidade de Lisboa -- UL,
Avenida Rovisco Pais 1, 1049 Lisboa, Portugal}
\begin{abstract}
Black holes possess trapping regions which lead to intriguing dynamical effects. By properly scattering test fields off a black hole, one can extract energy from it, leading to the growth of the amplitude of the test field in expense of the black hole's energy. Such a dynamical phenomenon is called superradiance.  Here, we study charged-scalar-field instabilities of Reissner-Nordstr\"{o}m black holes immersed in a $d-$dimensional de Sitter Universe. By performing a thorough frequency-domain analysis we compute the unstable quasinormal resonances and link their presence with a novel family of quasinormal modes associated with the existence and timescale of the cosmological horizon of pure de Sitter spacetime. Our results indicate that such an instability is caused by superradiance, while the increment of dimensions amplifies the growth rate and enlarges the region of the parameter space where, both massless and massive, test fields are unstable.
\end{abstract}

\maketitle
%

%%%%%%%%%%%%%%%%%%%%%%%%%%%%%%%%%%%%%%%%%%%%%%%%%%
\noindent{\bf{\em I. Introduction.}}
%%%%%%%%%%%%%%%%%%%%%%%%%%%%%%%%%%%%%%%%%%%%%%%%%%
The study of linear perturbations has a long  history in General Relativity (GR).  A perturbative analysis of black hole (BH) spacetimes was pioneered by Regge and Wheeler \cite{Regge}, and has proven crucial  in  several  contexts, ranging  from  astro-physics to high-energy physics \cite{Barack:2018yly,Pani:2013pma}.  

The stability analysis of BH spacetimes, an understanding of ringdown signals in the post-merger phase of a binary coalescence and their use in tests of GR, or even the analysis of fundamental light fields in the vicinities of BHs are some noteworthy examples where BH perturbation theory plays an important role \cite{Kokkotas:1999bd,Berti:2009kk,Konoplya:2011qq}.

Perturbing a BH with small fluctuations could lead to two possible outcomes; the BH is stable under perturbations, due to damping mechanisms that act on the BH exterior, and will relax after the initial disruption or the BH is unstable under perturbations and will inevitably disappear or evolve to another stable object. Although astrophysical BHs are expected to be stable under small fluctuations, a lot of concern has been given to BH solutions that might be prone to instabilities due to new phenomena that might be possibly unveiled. 

Quiet strikingly, it has been shown that one can extract energy from BHs through scattering techniques \cite{Penrose:1971uk,Bekenstein:1973mi} by properly probing them with test fields, and under certain circumstances the test fields can grow in time in expense of the BH's energy. This effect is called superradiance \cite{Brito:2015oca} and was explored in the context of rotating and charged BHs \cite{Detweiler,Vitor1,Vitor2,Vitor3,Vitor4,Vitor5,Kerr1,Kerr2,Kerr3,Kerr4,Kerr5,RN1,RN2,RN3,RN4,RN5,RN6,KN1,KN2}, stars \cite{Vicente:2018mxl,stars1,stars2} and other compact objects \cite{Maggio:2018ivz}.

Perturbation theory has revealed that BHs vibrate in a well described manner, exhibiting a discrete spectrum of preferable oscillatory modes, called quasinormal modes (QNMs) \cite{Chandrasekhar:1975zza,PhysRevLett.52.1361,Kokkotas:1999bd,Berti:2009kk,Konoplya:2011qq}. Linear perturbation theory has been an active field of study for decades and has been proven a very practical tool for testing the modal stability of BHs and compact objects, both analytically and numerically. Various studies have brought to light spacetimes which upon perturbations become unstable (for an incomplete list see \cite{Furuhashi:2004jk,Konoplya:2008au,Dolan:2012yt,Destounis:2018utr,Zhu:2014sya,Konoplya:2014lha,Dias:2010eu,Cardoso:2010rz,Konoplya:2013sba,Herdeiro:2013pia,Degollado:2013bha,Sanchis-Gual:2015lje,Li:2012nd,Li:2012rx,Li:2014fna,Li:2014gfg,Li:2014xxa,Li:2015mqa}).

An interesting study \cite{Konoplya:2008au} suggests that higher-dimensional Reissner-Nordstr\"om-de Sitter (RNdS) spacetimes are gravitationally unstable. Such a gravitational instability has been further examined in \cite{Cardoso:2010rz,Konoplya:2013sba,Tanabe:2015isb}. Specifically, $d-$dimensional RNdS BHs with $d>6$ and large enough mass and charge, are unstable under gravitational perturbations. Why only $d=4,5$ and $6-$dimensional RNdS BHs are favorable to be gravitationally stable is still unknown. 
%Due to the consideration of charge in such BHs, one could argue that for them to form through gravitational collapse, charged fields should be present.

More recently, a novel instability was found in $4-$dimensional RNdS BHs \cite{Zhu:2014sya,Konoplya:2014lha}. The $l=0$ charged scalar perturbation was proven to be unstable for various regions of the parameter space of RNdS BHs. The addition of an arbitrarily small amount of mass acts as a stabilization factor, as well as the increment of the scalar field charge beyond a critical value. Such an instability is caused by superradiance.

In this work, we investigate such an instability in higher-dimensions by employing a frequency-domain analysis. We analyze both massless and massive charged scalar perturbations in subextremal $d-$dimensional RNdS spacetime and show that the instability still persists. For simplicity, we narrow down our study in $d=4,5$ and $6$ dimensions. 

Intriguingly, we will show that the superradiant instability originates from a new family of QNMs which exists only in asymptotically de Sitter (dS) BHs and can be very well approximated by the QNMs of pure $d-$dimensional dS space. This novel family was very recently identified in asymptotically dS BHs for both scalar \cite{Jansen:2017oag,Cardoso:2017soq,PhysRevD.98.104007,Liu:2019lon} and fermionic perturbations \cite{Destounis:2018qnb}. Finally, we will demonstrate that as the spacetime dimensions increase, the instability is amplified, occurs for a larger region of the subextremal parameter space and still satisfies the superradiant condition. 

%%%%%%%%%%%%%%%%%%%%%%%%%%%%%%%%%%%%%%%%%%%%%%%%%%
\noindent{\bf{\em II. Charged scalar fields in higher-dimensional Reissner-Nordstr\"om-de Sitter spacetime.}}
%%%%%%%%%%%%%%%%%%%%%%%%%%%%%%%%%%%%%%%%%%%%%%%%%%
The $d-$dimensional RNdS spacetime is described by the metric 
\begin{equation}
ds^2=-f(r)dt^2+\frac{1}{f(r)}dr^2+r^2d\Omega^2_{d-2},
\end{equation}
with 
\begin{equation*}
d\Omega^2_{d-2}=d\chi_2^2+\prod_{i=2}^{d-2}\sin^2\chi_i d\chi^2_{i+1}.
\end{equation*}
The metric function reads
\begin{align}
%\nonumber
 f(r)&=1-\frac{m}{r^{d-3}}+\frac{q_0}{r^{2(d-3)}}-\frac{2\Lambda }{(d-2)(d-1)}r^2,
% &=\frac{\Lambda}{3r^2}\left(r-r_0\right)(r-r_-)(r-r_+)(r_c-r),
\end{align}
where $\Lambda$ is the cosmological constant and $m$, $q_0$ are functions related to the ADM mass $M$ and electric charge $Q$ of the BH, respectively,
\begin{equation}
M=\frac{d-2}{16\pi}w_{d-2}m,\,\,Q=\frac{\sqrt{2(d-2)(d-3)}}{8\pi}w_{d-2}q_0,
\end{equation}
with $w_d=2\pi^{\frac{d+1}{2}}/\Gamma(\frac{d+1}{2}),$ the volume of the unit $d-$sphere.
The causal structure of a subextremal higher-dimensional RNdS BH possesses three distinct horizons, namely the Cauchy $r=r_-$, event $r=r_+$ and cosmological horizon $r=r_c$, where $r_-<r_+<r_c$. The associated electromagnetic potential is
\begin{equation}
A=-\sqrt{\frac{d-2}{2(d-3)}}\frac{q_0}{r^{d-3}}dt.
\end{equation}
The propagation of a massive charged scalar field on a fixed $d-$dimensional RNdS background is governed by the Klein-Gordon equation 
\begin{equation}
(D^\nu D_\nu-\mu^2)\psi=0, 
\end{equation}
where $D_\nu=\nabla_\nu-iqA_\nu$ is the covariant derivative and $\mu$, $q$ are the mass and charge of the field, respectively. By expanding $\psi$ in terms of spherical harmonics with harmonic time dependence,
\begin{equation}
\label{expand}
\psi=\sum_{lm}\frac{\Psi_{l m}(r)}{r^{\frac{d-2}{2}}}Y_{lm}(\chi)e^{-i\omega t},
\end{equation}
and dropping the subscripts on the radial functions, we obtain the master equation
\begin{equation}
\label{master_eq_RNdS}
\frac{d^2 \Psi}{d r_*^2}+\left[\omega^2-2\omega\Phi(r)-V(r)\right]\Psi=0\,,
\end{equation}
where 
\begin{equation}
\Phi(r)=\frac{q_0 q}{\sqrt{\frac{2(d-3)}{d-2}}r^{d-3}},
\end{equation}
is the electrostatic potential, $dr_*=dr/f(r)$ is the tortoise coordinate and 
\begin{align}
\nonumber
V(r)=f(r)\left(\mu^2+\frac{l(l+d-3)}{r^2}+\frac{(d-2)f^\prime(r)}{2r}\right.\\\label{RNdS_general potential}
\left.+f(r)\frac{(d-4)(d-2)}{4r^2}\right)-\Phi(r)^2,
\end{align}
is the effective potential, with $l$ an angular number, corresponding to the eigenvalue of the spherical harmonics. Prime denotes the derivative with respect to the radial coordinate $r$. 

We are interested in the characteristic QNMs $\omega$ of such spacetime, obtained by imposing the boundary conditions~\cite{Berti:2009kk}
\begin{equation}
\label{bcs}
\Psi \sim
\left\{
\begin{array}{lcl}
e^{-i (\omega-\Phi(r_+))r_* },\,\,\,\quad r \rightarrow r_+, \\
&
&
\\
 e^{+i(\omega-\Phi(r_c))r_*},\,\,\,\,\quad r \rightarrow r_c.
\end{array}
\right.
\end{equation}
The QN frequencies are characterized, for each $l$, by an integer $n\geq 0$ labeling the mode number. The fundamental mode $n=0$ corresponds, by definition, to the non-vanishing frequency with the smallest (by absolute value) imaginary part. 

According to the time dependence in \eqref{expand}, when $\text{Im}(\omega)<0$, the perturbation will decay in time and we have a stable mode. On the other hand, if $\text{Im}(\omega)>0$, then the perturbation will grow in time and we have an unstable mode. Due to the underlying symmetry of (\ref{master_eq_RNdS}) $\text{Re}(\omega)\rightarrow-\text{Re}(\omega)$ and $\Phi(r)\rightarrow-\Phi(r)$, we will consider only cases where $qQ>0$.
\begin{figure*}[t!]
\includegraphics[scale=0.25]{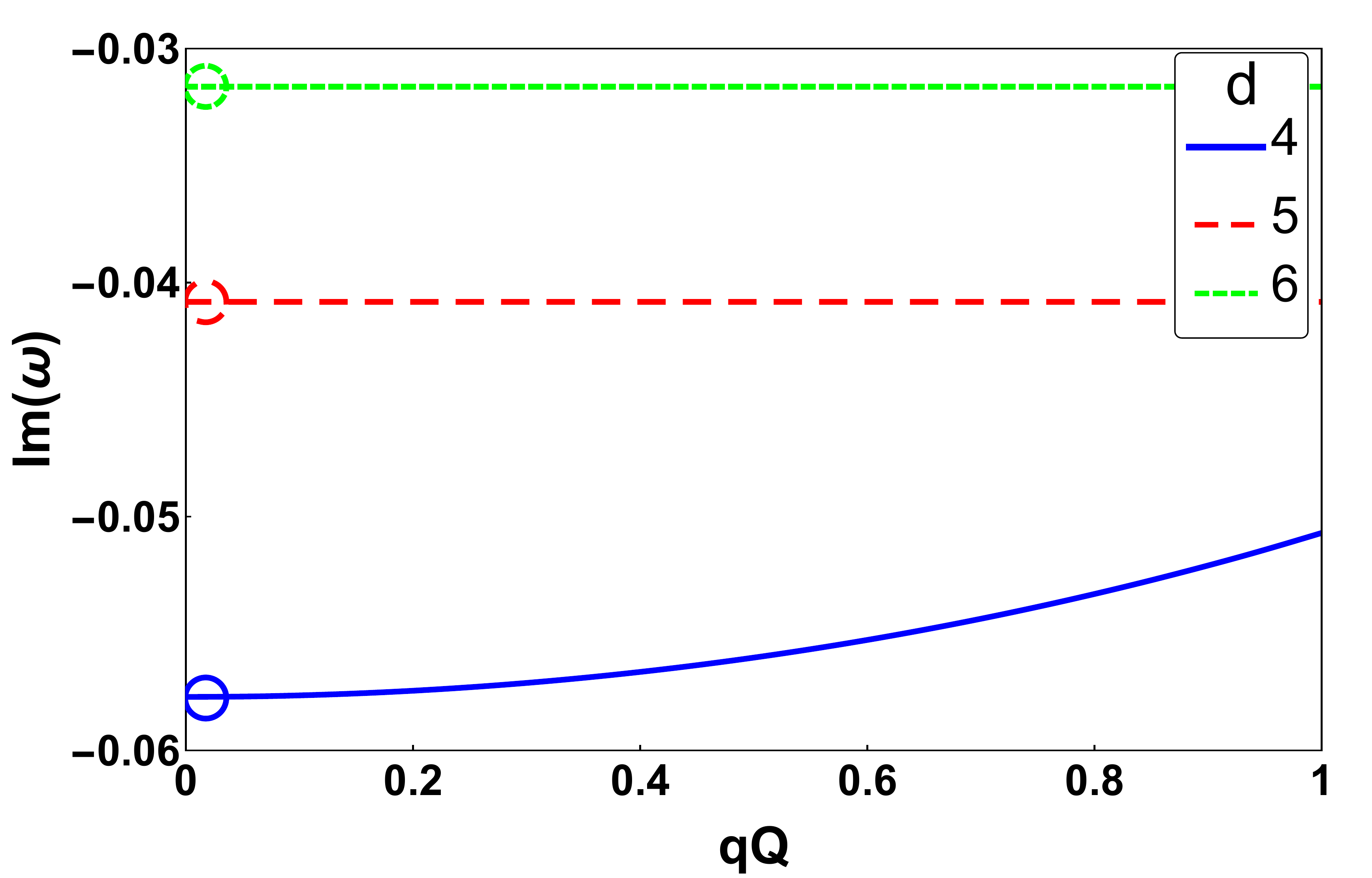}
\includegraphics[scale=0.25]{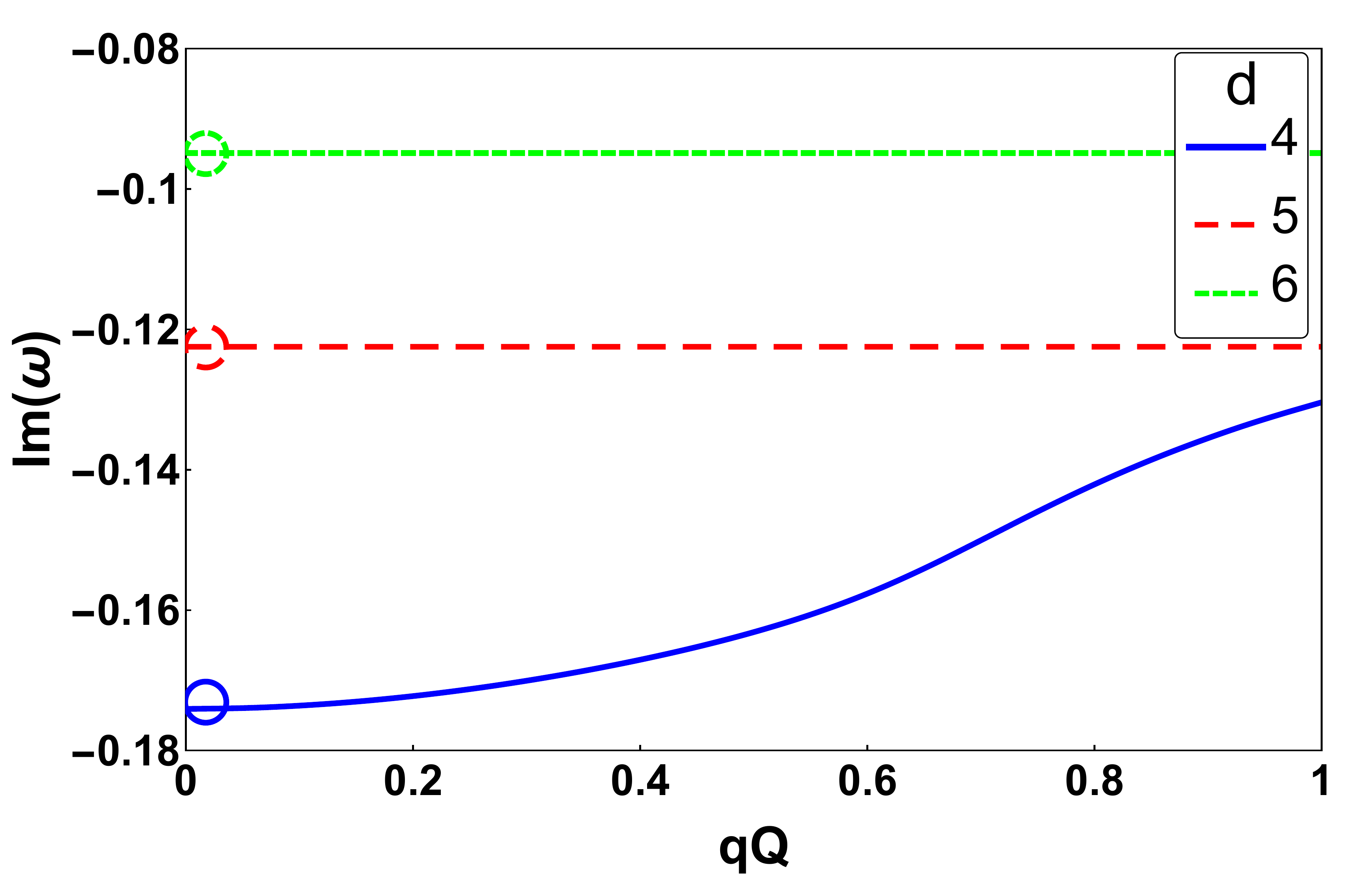}
\caption{Imaginary parts of $n=0$ (left) and $n=1$ (right) BH dS QNMs with $l=1$ of a $d-$dimensional RNdS BH with $M=1$, $\Lambda=0.01$ and $Q=0.5$ versus the charge coupling $qQ$. The circles designate the respective scalar modes of pure dS space with $\Lambda=0.01$.}
\label{dS}
\end{figure*}

The results shown in the following sections were obtained mostly with the Mathematica package of \cite{Jansen:2017oag} (based on methods developed in \cite{Dias:2010eu}), and checked in various cases with a Wentzel–Kramers–Brillouin (WKB) approximation \cite{Iyer:1986np} and a matrix method that was developed based on the non-grid based interpolation scheme proposed in \cite{KaiLin1}.

%%%%%%%%%%%%%%%%%%%%%%%%%%%%%%%%%%%%%%%%%%%%%%%%%%
\noindent{\bf{\em III. Superradiance in higher-dimensional Reissner-Nordstr\"om-de Sitter spacetime.}}
%%%%%%%%%%%%%%%%%%%%%%%%%%%%%%%%%%%%%%%%%%%%%%%%%%
The effect of superradiantly unstable charged scalar fields scattering off a $4-$dimensional RNdS spacetime has been analyzed in \cite{Zhu:2014sya,Konoplya:2014lha}. To generalize the study in $d-$dimensions, we consider charged massive scalar fields scattering off the BH effective potential. Such a scenario demands that we impose the following boundary conditions in (\ref{master_eq_RNdS}):
\begin{equation}
\label{scat}
\Psi \sim
\left\{
\begin{array}{lcl}
B e^{-i (\omega-\Phi(r_+))r_* },\,\,\,\quad\quad\quad\quad\quad\quad\,\, r \rightarrow r_+, \\
&
&
\\
 e^{-i(\omega-\Phi(r_c))r_*} + A e^{i(\omega-\Phi(r_c))r_*},\,\,\, r \rightarrow r_c.
\end{array}
\right.
\end{equation}
Eq. \eqref{master_eq_RNdS} also possesses a complex conjugate solution $\bar{\Psi}$. It is easy to prove that the Wronskian of the independent solutions $\Psi$, $\bar{\Psi}$ is $r_*$-independent and therefore, conserved. Due to the conservation of the Wronskian, we can derive the following relation associating the reflection ($A$) and transmission ($B$) coefficients as follows:
\begin{equation}
\label{relation}
|A|^2=1-\frac{\omega-\Phi(r_+)}{\omega-\Phi(r_c)}|B|^2.
\end{equation}
From \eqref{relation} we recognize that superradiance occurs if
\begin{equation}
\label{suprad}
\Phi(r_c)<\omega<\Phi(r_+),
\end{equation}
which designates that the amplitude of the reflected wave is larger than the amplitude of the incident wave. In \cite{Konoplya:2014lha} it was proven that the real part of $\omega$ satisfying (\ref{suprad}) is the necessary, but not sufficient, condition for the instability. Thus, the instability can only take place when (\ref{suprad}) is satisfied, but on the other hand, (\ref{suprad}) can also be satisfied by stable modes. As we will see in the following, the same holds for higher-dimensional RNdS spacetimes. This result is qualitatively different compared to the higher-dimensional asymptotically anti-de Sitter BHs, where the necessary condition is also the sufficient one \cite{Kodama:2009rq, Wang:2014eha}.

%%%%%%%%%%%%%%%%%%%%%%%%%%%%%%%%%%%%%%%%%%%%%%%%%%
\noindent{\bf{\em IV. The modes of asymptotically de Sitter black holes.}}
%%%%%%%%%%%%%%%%%%%%%%%%%%%%%%%%%%%%%%%%%%%%%%%%%%
The QNMs of asymptotically flat BHs are very well analyzed. Many spherically-symmetric BHs are known to possess a region outside the event horizon which completely characterizes the QN spectrum of perturbations with large frequencies. This region is called the photon-sphere (PS), i.e. a hypersurface where null particles are trapped in unstable circular orbits. It has been shown \cite{Cardoso:2008bp} that such a region controls the stability and decay of perturbations on the exterior of many BHs\footnote{There are some counter-examples of this concept in Einstein-Lovelock gravity \cite{Konoplya:2017wot} as well as near-extremal or extremal BHs \cite{Hod:2013eea,Hod:2017gvn,Cardoso:2003sw,Cardoso:2017soq}.}. For example, the decay timescale of perturbations is associated with the instability timescale of null geodesics at the PS.

\begin{figure*}[ht!]
\subfigure{\includegraphics[scale=0.17]{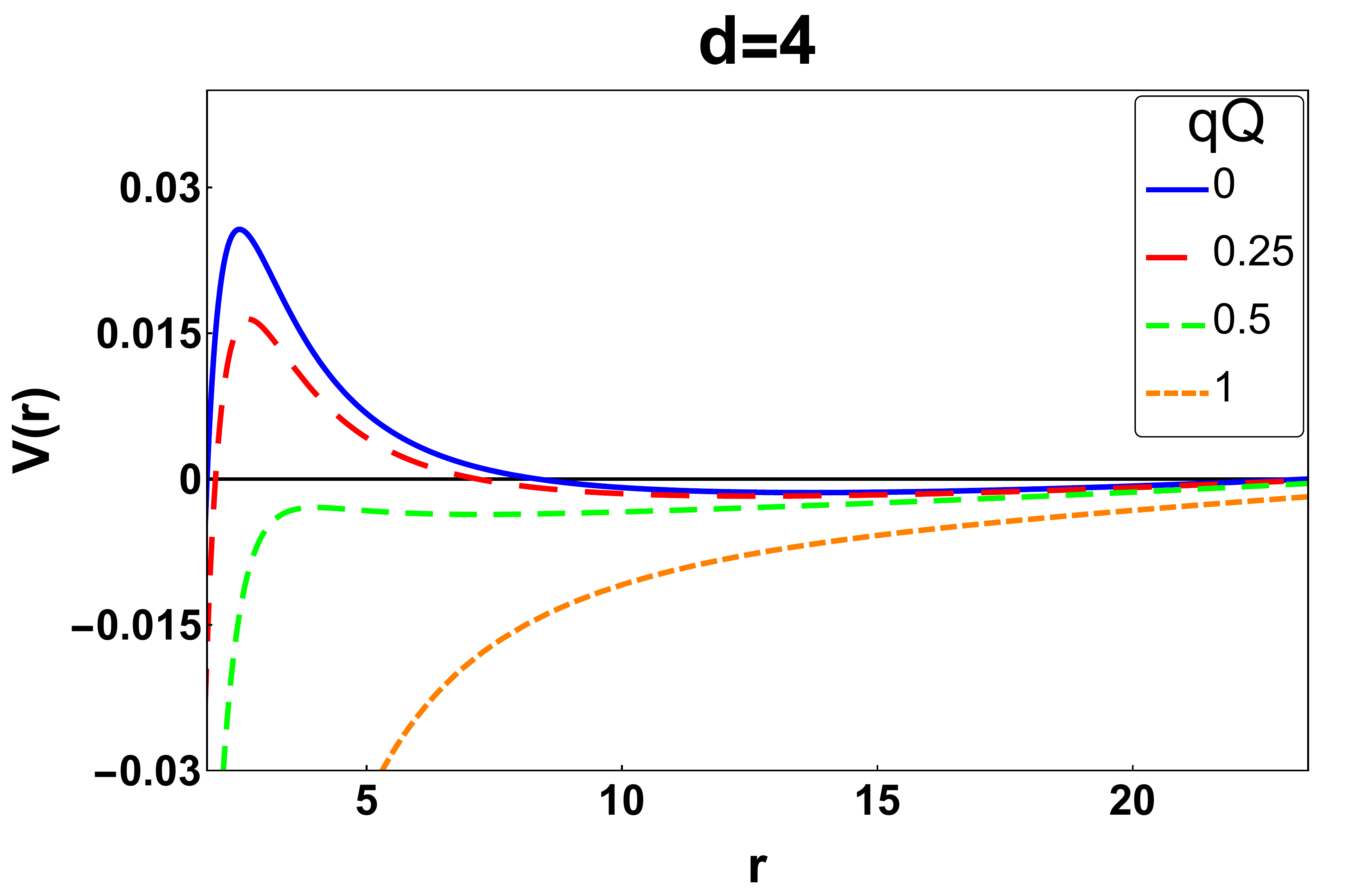}}\hskip -1ex
\subfigure{\includegraphics[scale=0.17]{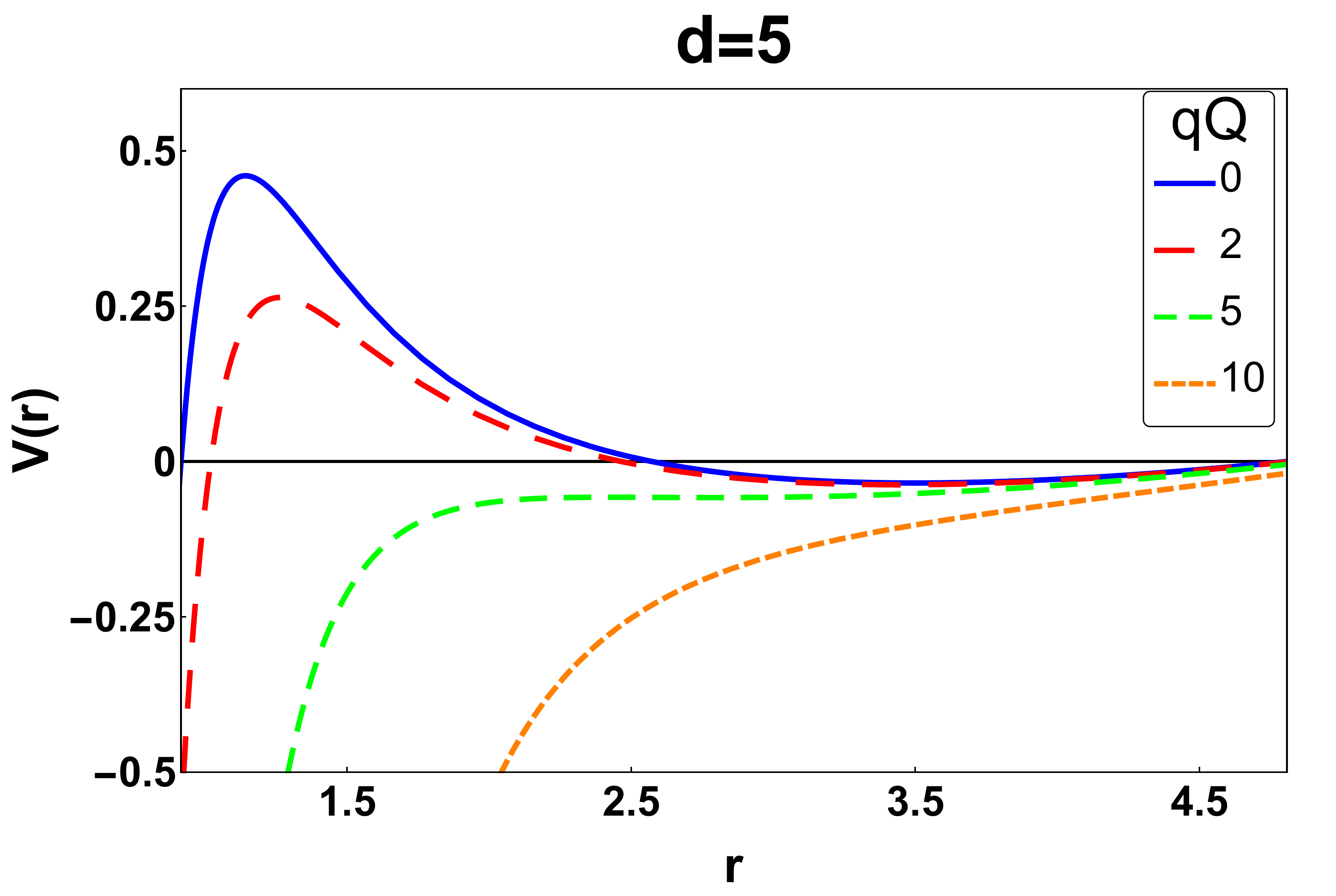}}\hskip -1ex
\subfigure{\includegraphics[scale=0.17]{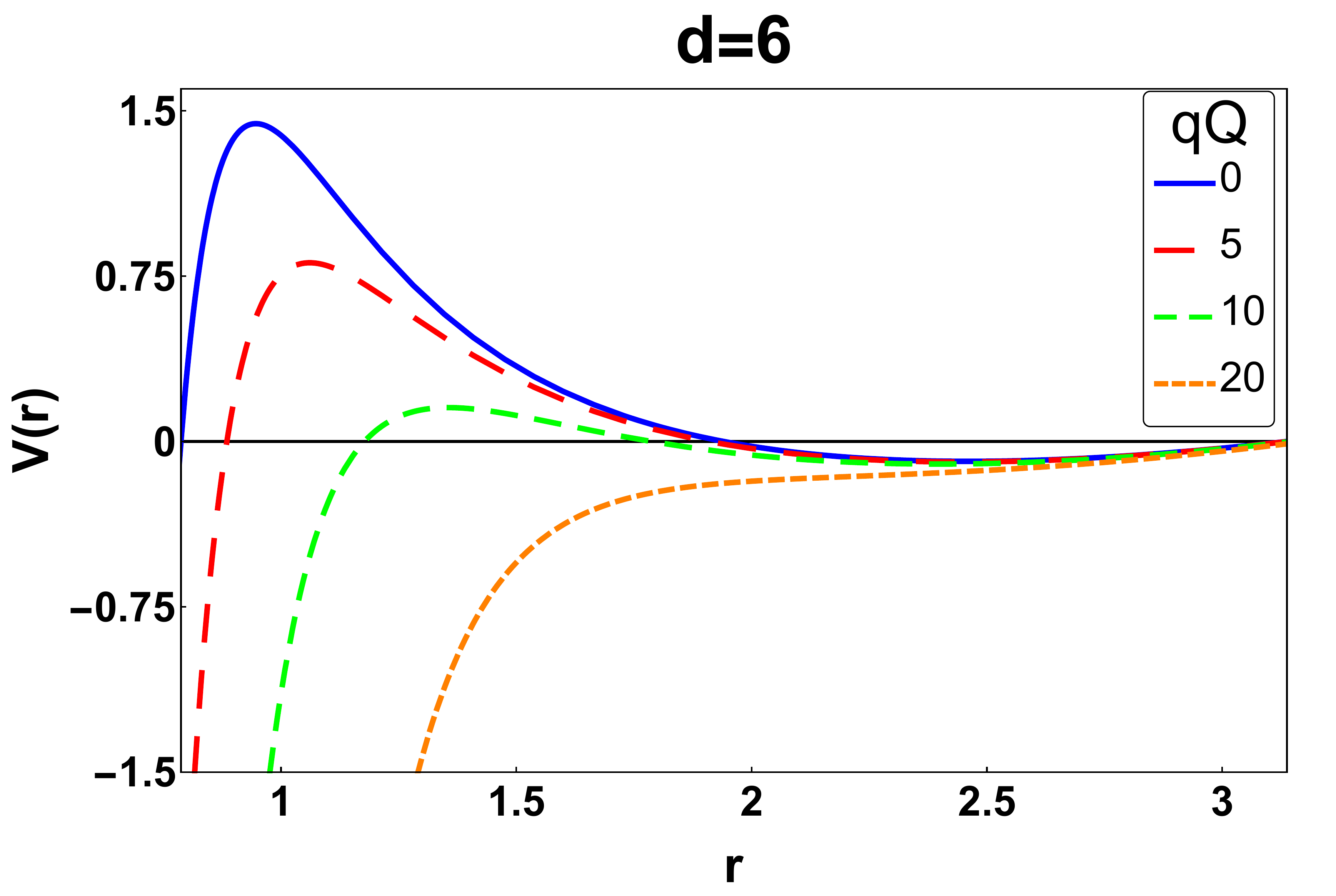}}\vskip -3ex
\subfigure{\includegraphics[scale=0.17]{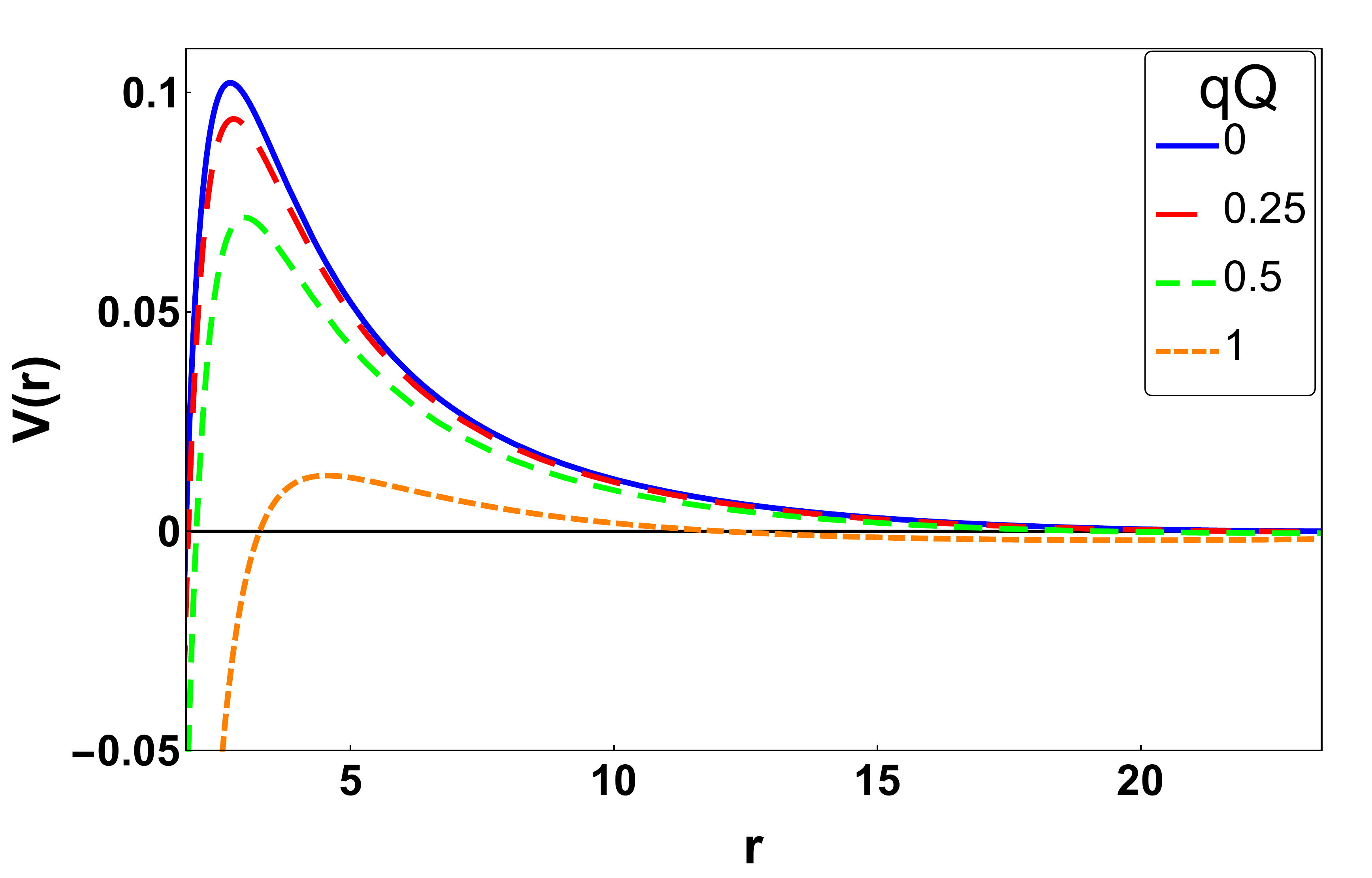}}\hskip -1ex
\subfigure{\includegraphics[scale=0.17]{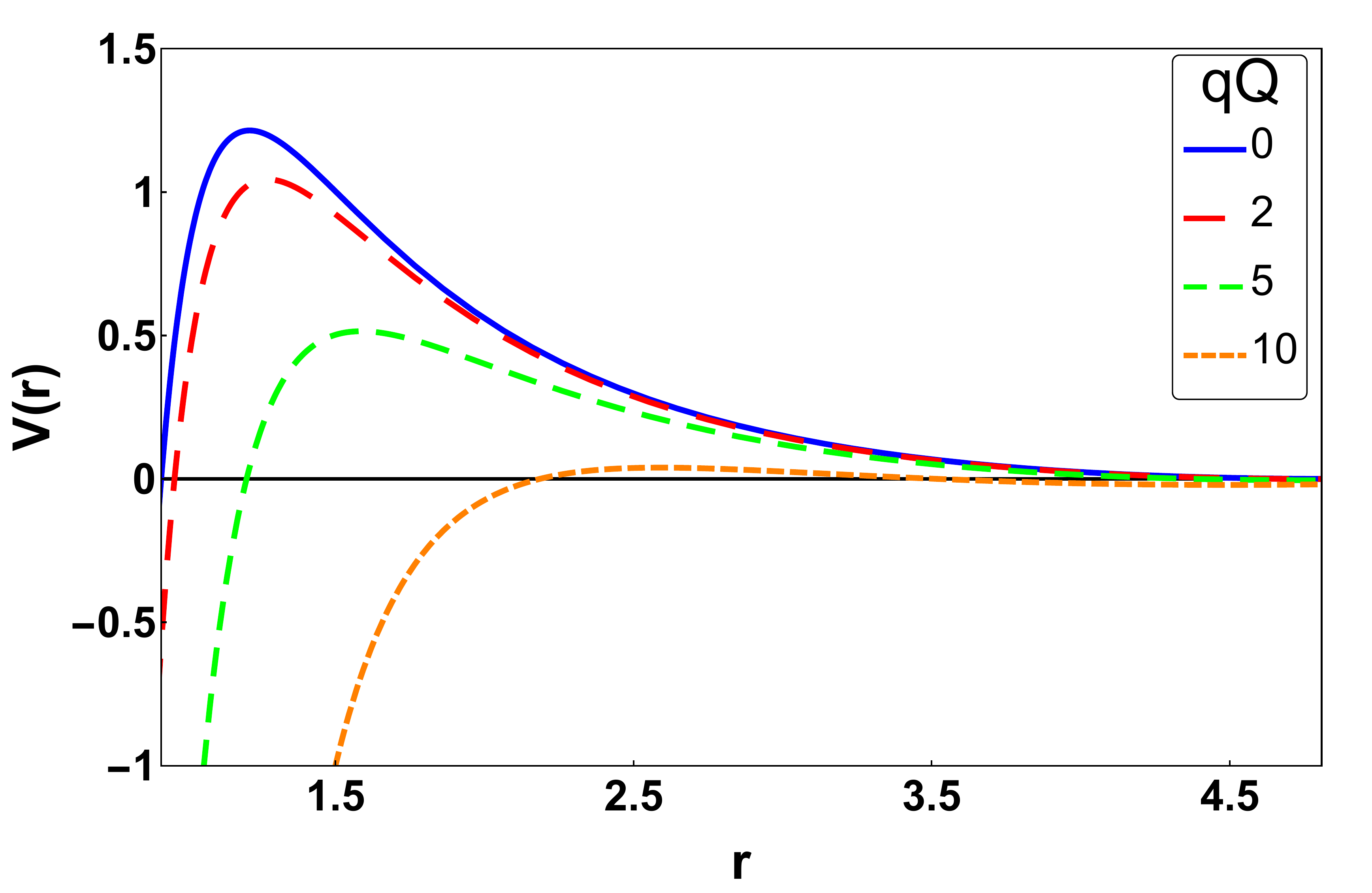}}\hskip -1ex
\subfigure{\includegraphics[scale=0.17]{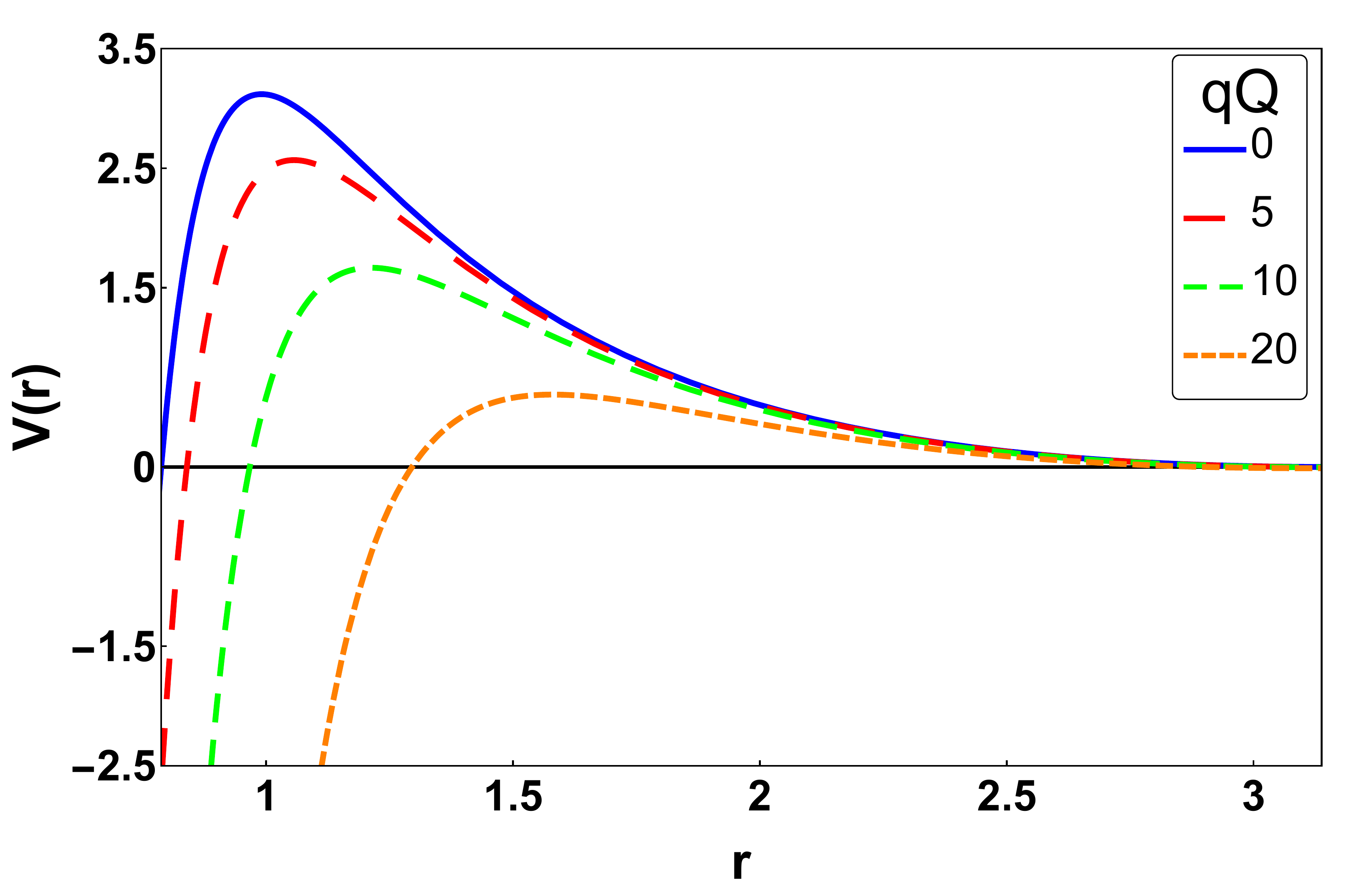}}
\caption{Effective potentials of $l=0$ (top panel) and $l=1$ (bottom panel) massless charged scalar perturbations of a $d-$dimensional RNdS BH with $M=1$ and $Q=0.5$ for various charge couplings $qQ$. The cosmological constants used are $\Lambda=0.005$ for $d=4$, $\Lambda=0.25$ for $d=5$ and $\Lambda=1$ for $d=6$.}
\label{potentials}
\end{figure*}
A complete picture of the modes of asymptotically dS BHs was lacking till recently. Besides the PS modes, there is another family which is completely distinct from the former and is associated with the accelerated expansion of the Universe \cite{Brill:1993tw,Rendall:2003ks}. The BH dS family of modes does not converge in any limit to the PS family and it was first found in \cite{Jansen:2017oag,Cardoso:2017soq} for Schwarzschild-dS (SdS) and RNdS BHs. Such modes have a surprisingly weak dependence of the electric charge of the BH, if it has one, and can be very well approximated by the modes of purely de Sitter space. The scalar QNMs of pure $d-$dimensional de Sitter space are \cite{LopezOrtega:2006my,VasydS}
\begin{eqnarray}
\label{family1}
\omega_{\rm pure \,\rm dS}/\kappa_c^{\rm dS} &=&-i (l+2n)\,,\\
\omega_{\rm pure \,\rm dS}/\kappa_c^{\rm dS} &=&-i(l+2n+d-1)\,,\label{family2}
\end{eqnarray}
where $\kappa_c^\text{dS}=\sqrt{2\Lambda/(d-2)(d-1)}$ is the surface gravity of the cosmological horizon of pure $d-$dimensional dS space. 
The fundamental $n=0$ BH dS QNMs reported in \cite{Jansen:2017oag,Cardoso:2017soq,Liu:2019lon} are approximated, with high accuracy, by \eqref{family1}, while higher overtones have larger deformations to \eqref{family1} and \eqref{family2}. Equivalent results have been obtained for fermionic perturbations on RNdS spacetime \cite{Destounis:2018qnb}. 

In Fig. \ref{dS} we show the imaginary parts of an $l=1$ BH dS mode for various dimensions with respect to the charge coupling $qQ$. We observe that the fundamental mode and the first overtone of this family matches very well Eq. \eqref{family1} (depicted with circles) for $qQ=0$. Eq. \eqref{family2} only contributes for higher overtones. It seems that as the dimensions increase, the modes become more robust against the increment of $qQ$, at least for the given choice of parameters\footnote{But still, with proper increment of $qQ$, even the higher-dimensional BH dS QNMs are affected. For example, the fundamental BH dS mode for $d=5$, $M=1$, $\Lambda=0.01$ and $qQ=50$ is $\text{Im}(\omega)=-0.0363 i$, while for $qQ=0$ is $\text{Im}(\omega)=-0.0408 i$. The fundamental BH dS mode for $d=6$, $M=1$, $\Lambda=0.5$ and $qQ=50$ is $\text{Im}(\omega)=-0.2133 i$, while for $qQ=0$ is $\text{Im}(\omega)=-0.2236 i$.}. In any case, all $l>0$ modes belonging to the BH dS family are stable. This can be explained by the form of the effective potential of the fixed background. 

In Fig. \ref{potentials} we demonstrate that for $l=1$, \eqref{RNdS_general potential} does not form any potential wells outside the PS which might designate a possible long-lived mode or even an unstable one. On the contrary, for $l=0$ scalar perturbations, \eqref{RNdS_general potential} exhibits a potential well between the PS and the cosmological horizon (see Fig. \ref{potentials}). Here, we show that the potential wells found in \cite{Zhu:2014sya} continue to exist in higher dimensions which is an indication that instabilities might occurs even in $d>4$ RNdS BHs under charged scalar perturbations. 

\begin{figure*}[ht!]
\subfigure{\includegraphics[scale=0.17]{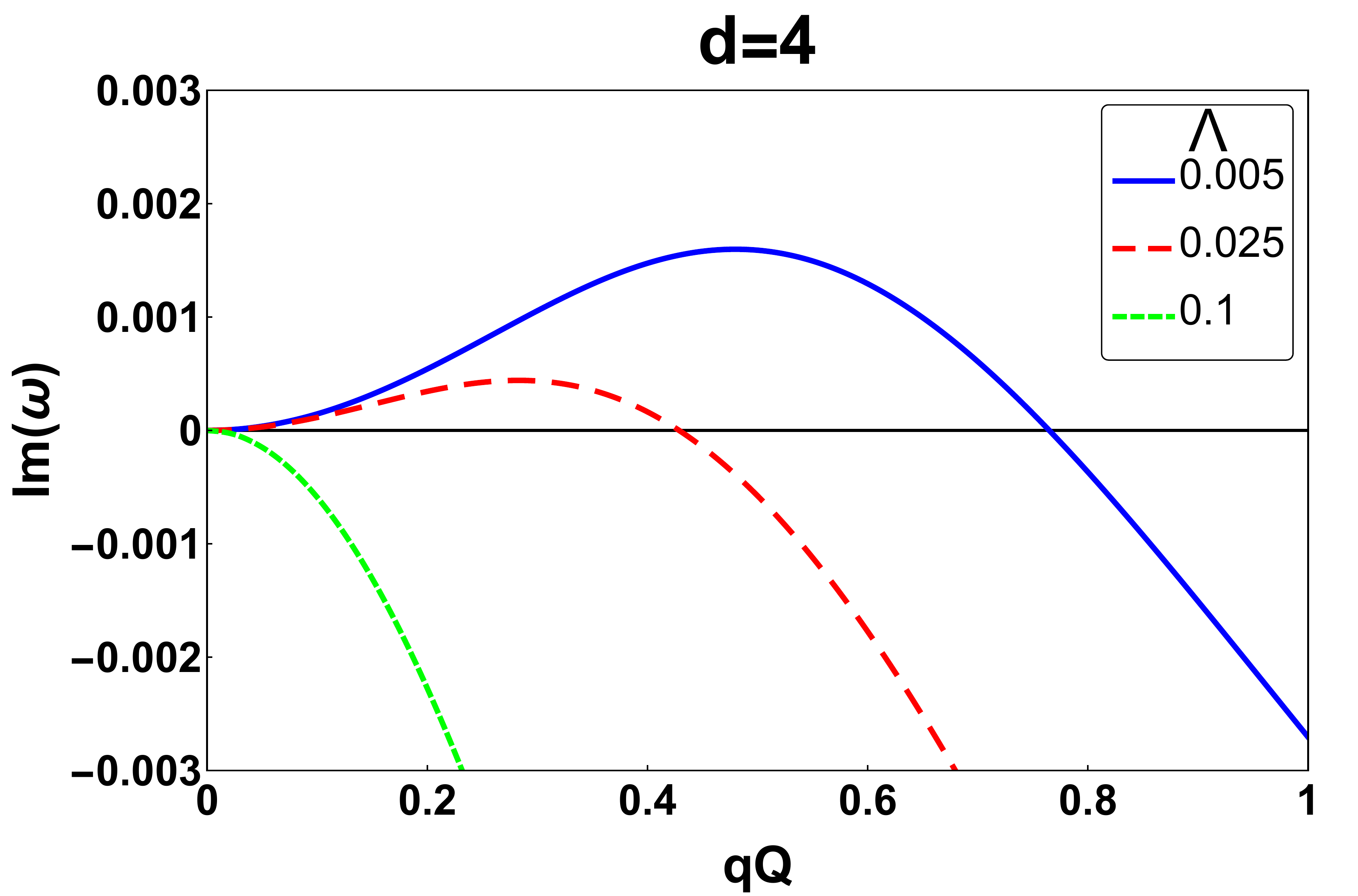}}
\hskip -2ex
\subfigure{\includegraphics[scale=0.17]{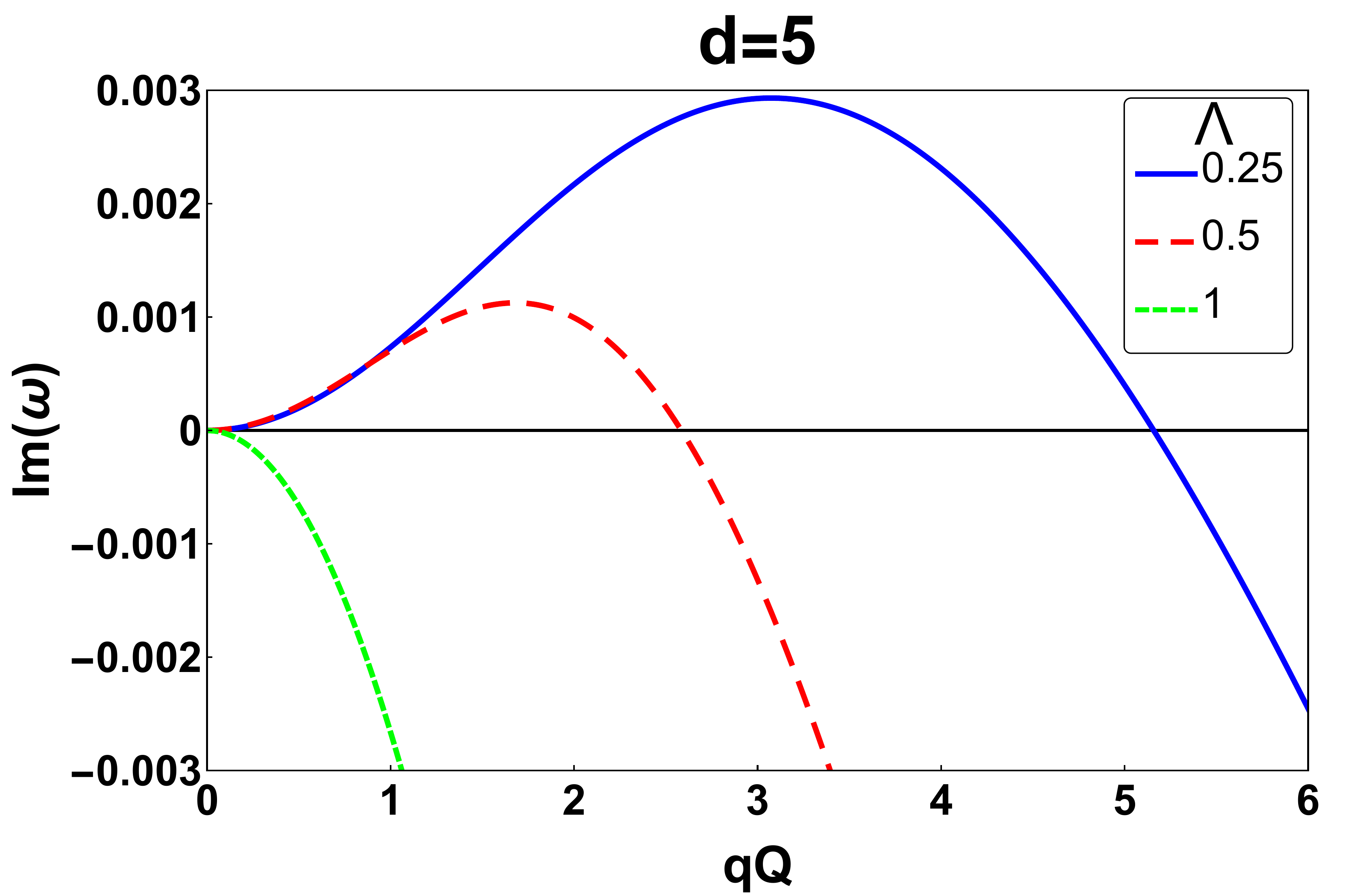}}
\hskip -2ex
\subfigure{\includegraphics[scale=0.17]{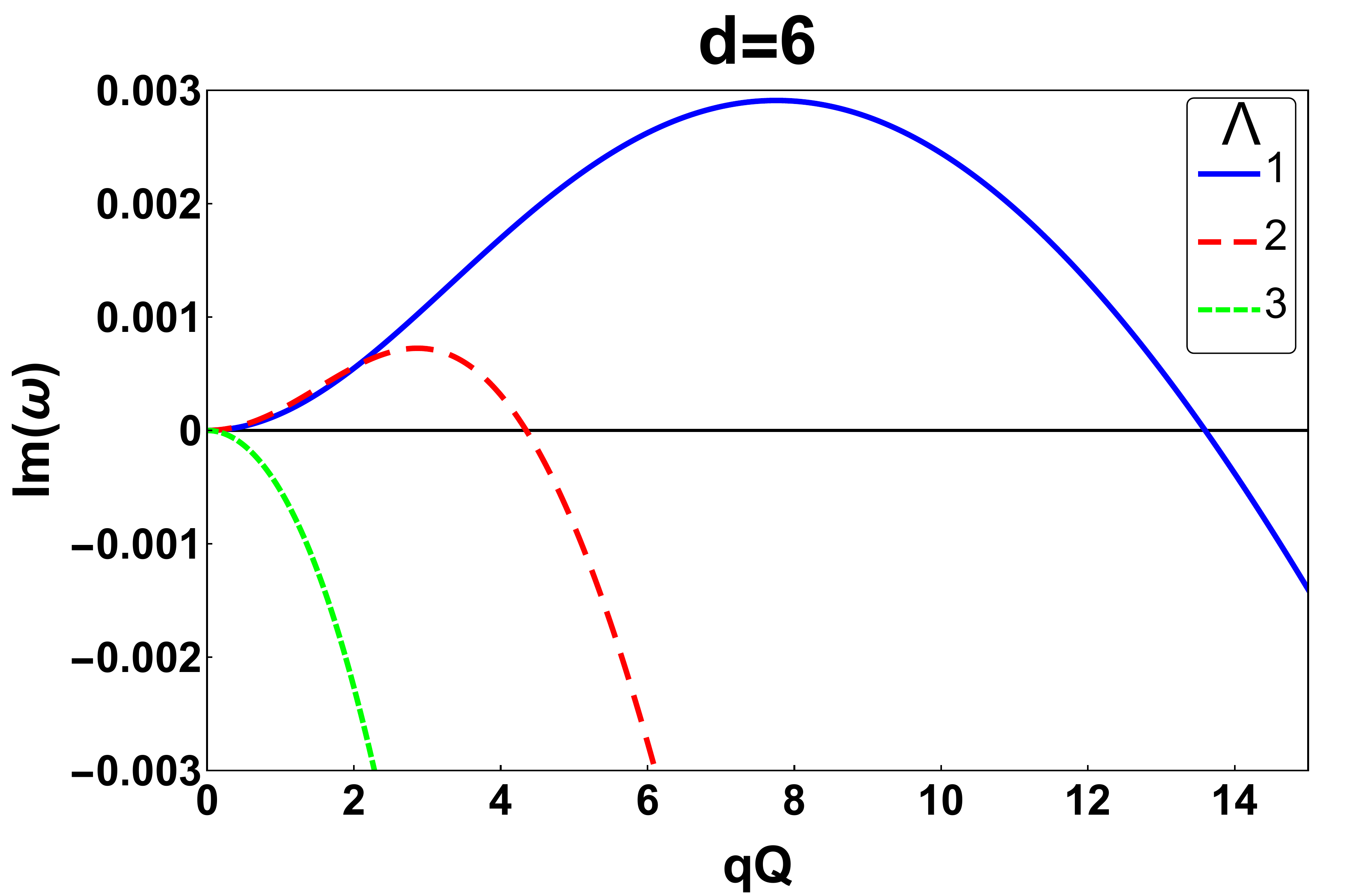}}
\vskip -3ex
\subfigure{\includegraphics[scale=0.17]{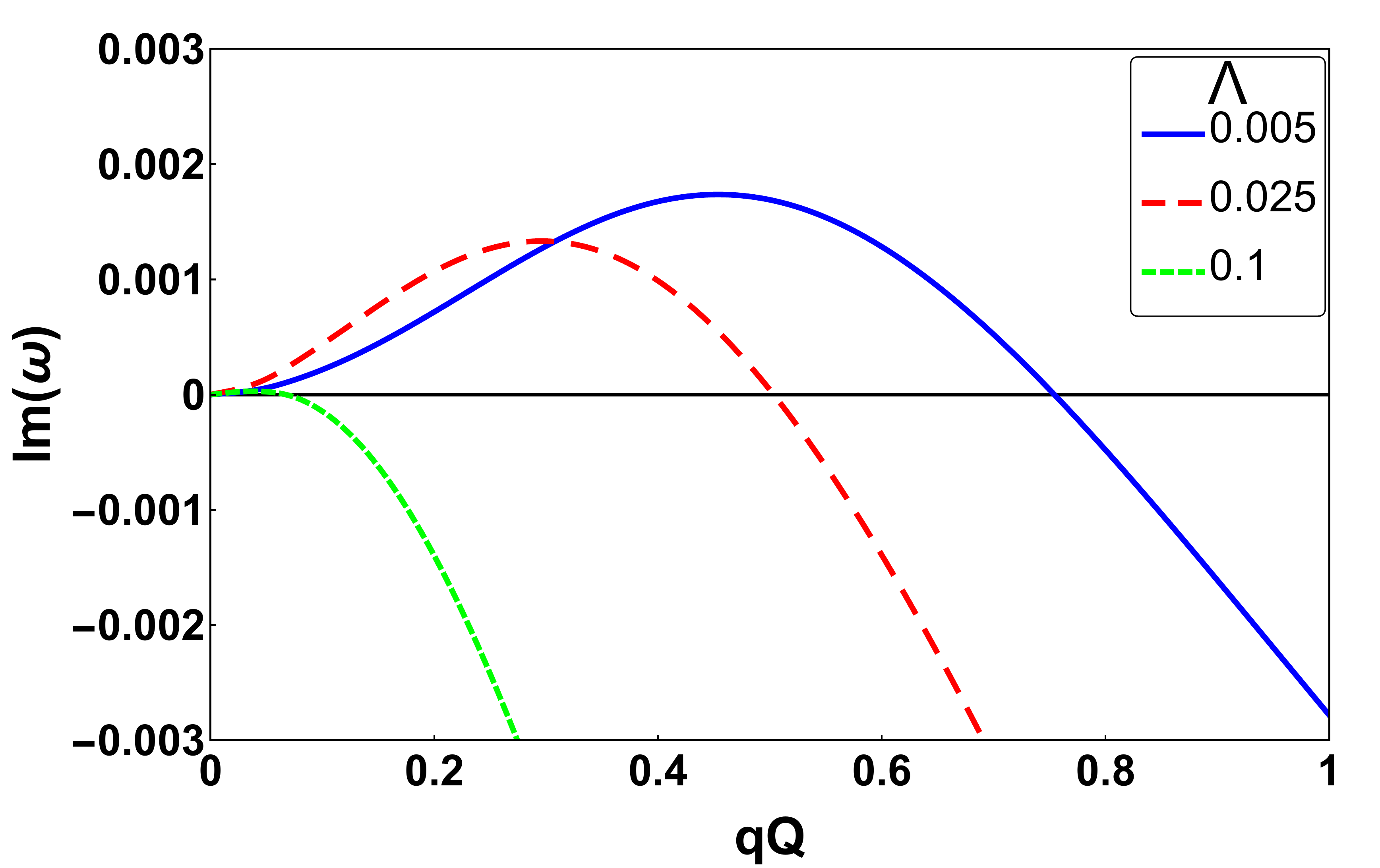}}
\hskip -2ex
\subfigure{\includegraphics[scale=0.17]{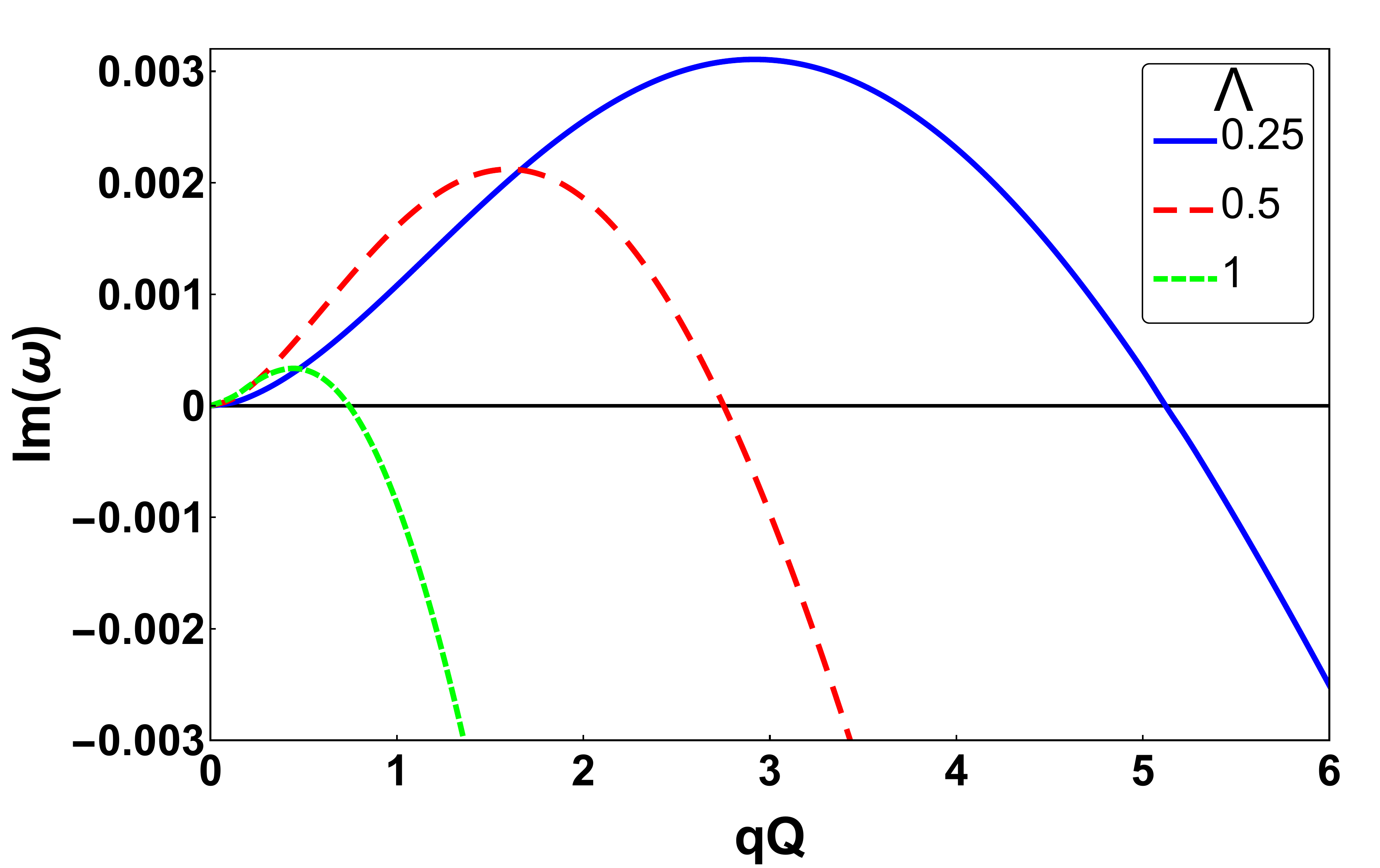}}
\hskip -2ex
\subfigure{\includegraphics[scale=0.17]{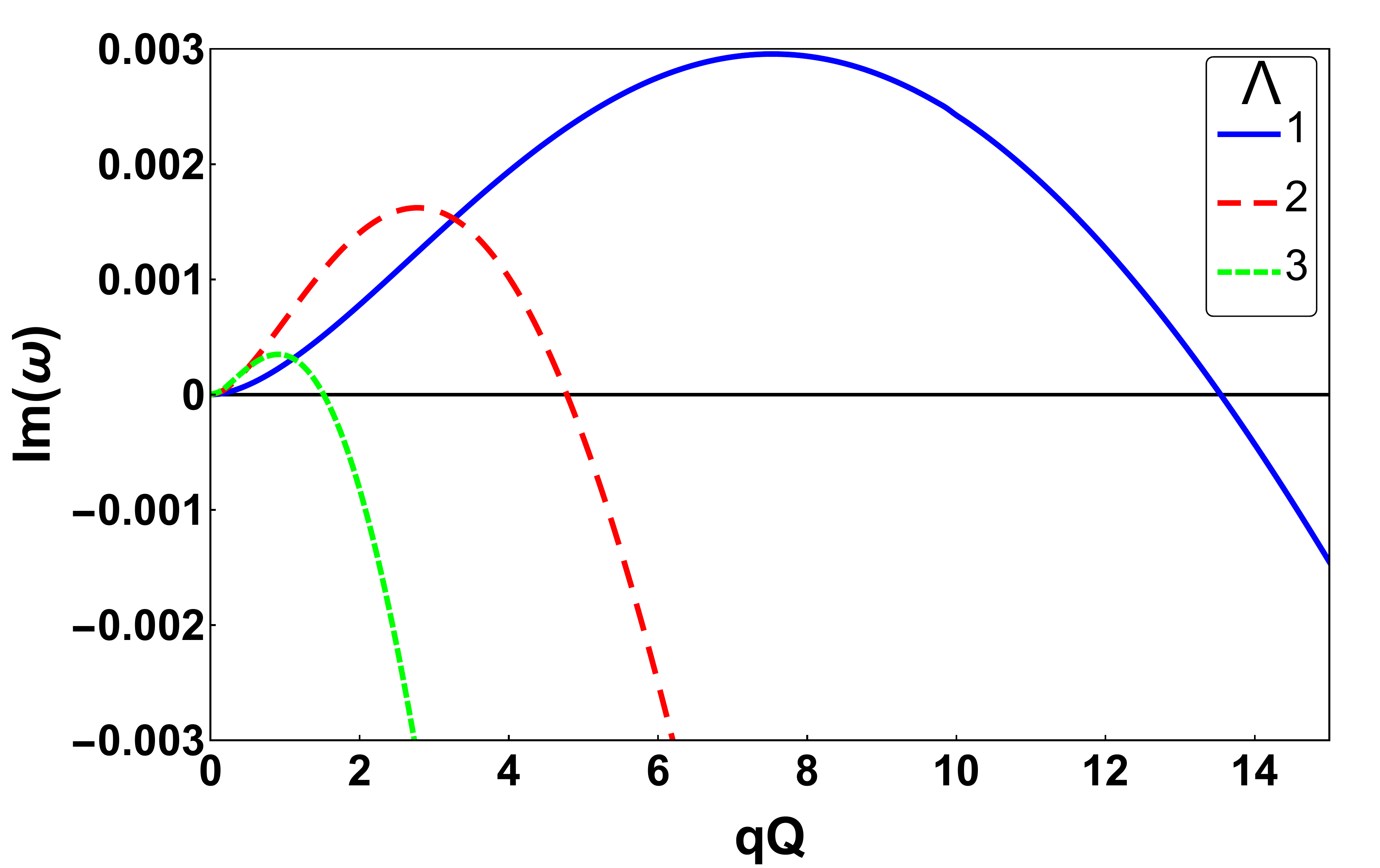}}
\caption{Imaginary parts of $l=0$ charged massless scalar perturbations on a fixed $d-$dimensional RNdS BH with $M=1$, $Q=0.5$ (top panel) and $Q=0.999\, Q_\text{max}$ (bottom panel) versus the charge coupling $qQ$, for various $\Lambda$.}
\label{massless}
\end{figure*}

It is crucial to note the existence of a "zero-mode" for $l=0$ scalar perturbations (see equation \eqref{family1}). Various time-domain calculations of linear perturbations in SdS and RNdS spacetimes have shown that neutral $l=0$ scalar perturbations do not decay in time but rather relax at a constant value \cite{Brady:1996za,Brady:1999wd,Zhu:2014sya,Konoplya:2014lha}, in contrast to $l>0$ fields which decay exponentially in time. This exactly occurs due to the existence of the "zero-mode" of the BH dS family. 

In the following sections, we will see that such modes are responsible for the superradiant instability of charged scalar fields in higher-dimensional RNdS spacetime.

%%%%%%%%%%%%%%%%%%%%%%%%%%%%%%%%%%%%%%%%%%%%%%%%%%
\noindent{\bf{\em V. Unstable massless charged scalar fields.}}
%%%%%%%%%%%%%%%%%%%%%%%%%%%%%%%%%%%%%%%%%%%%%%%%%%
In this section, we focus on the dominant modes of charged massless scalar fields in $d-$dimensional RNdS. The dominant modes will be the ones that control the dynamical evolution of the perturbation at late times. As shown in the previous section, unstable modes may occur for $l=0$ while $l>0$ are always stable.

We have identified that the existence of the "zero-mode" is directly linked to the QNMs of pure $d-$dimensional dS space and that this mode is prone to instabilities if charge is introduced to the perturbing field (see \cite{Zhu:2014sya,Konoplya:2014lha}). As shown in Fig. \ref{massless}, for $qQ> 0$ the "zero-mode" evolves to QNMs with positive imaginary parts, thus unstable. The increment of $qQ$ increases the imaginary part of the perturbation till its maximized. Beyond that point, the instability is saturated and the family acquires a negative imaginary part, thus restoring the modal stability. The stabilization due to large $qQ$ can be explained through the form of the effective potential. 

In Fig. \ref{potentials}, we demonstrate that the $4-$dimensional picture of \cite{Zhu:2014sya} remains similar in higher dimensions. By increasing $qQ$, (\ref{RNdS_general potential}) acquires a potential well which serves as a trapping region for waves to be captured and amplified. The instability occurs in this parametric region. More increment of $qQ$ leads to a purely negative effective potential without a potential well. 

The increment of dimensions, though, make the effective potential more robust to the increment of $qQ$. Hence, higher values of $qQ$ are required for the potential well to vanish. This can be explained by the fact that \eqref{RNdS_general potential} forms a deeper potential well as the dimensions increase. This leads to the amplification of the instability, as well as the enlargement of the parameter space region where the instability occurs. 

The real part of the unstable modes increases monotonously with respect to $qQ$ (see Table \ref{table1}). The imaginary part decreases more rapidly for larger $\Lambda$ and smaller $Q$. It is evident that as we approach extremality, the peak of instability is slightly increased.

\begin{figure*}[ht!]
\subfigure{\includegraphics[scale=0.17]{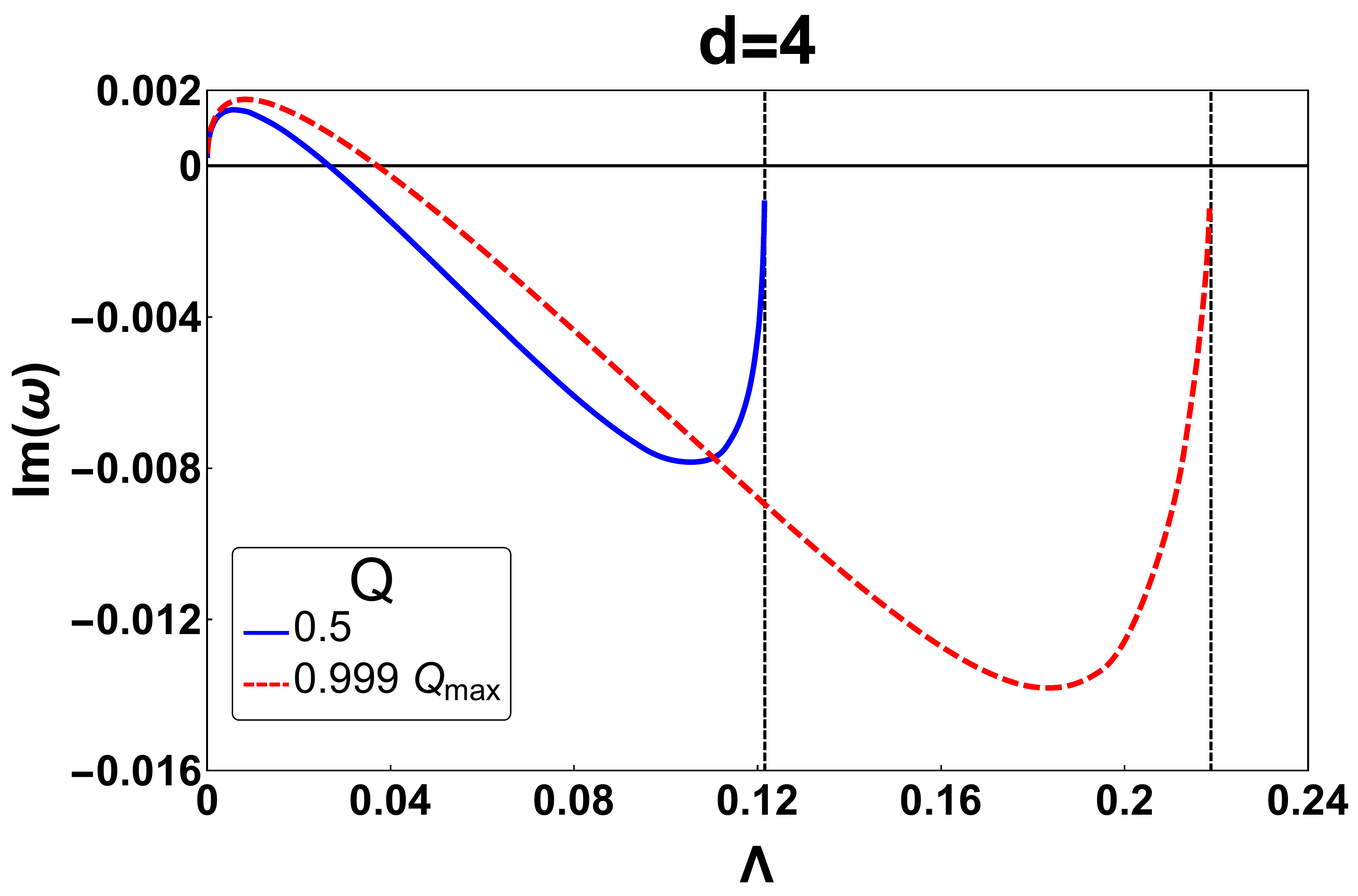}}
\hskip -2ex
\subfigure{\includegraphics[scale=0.17]{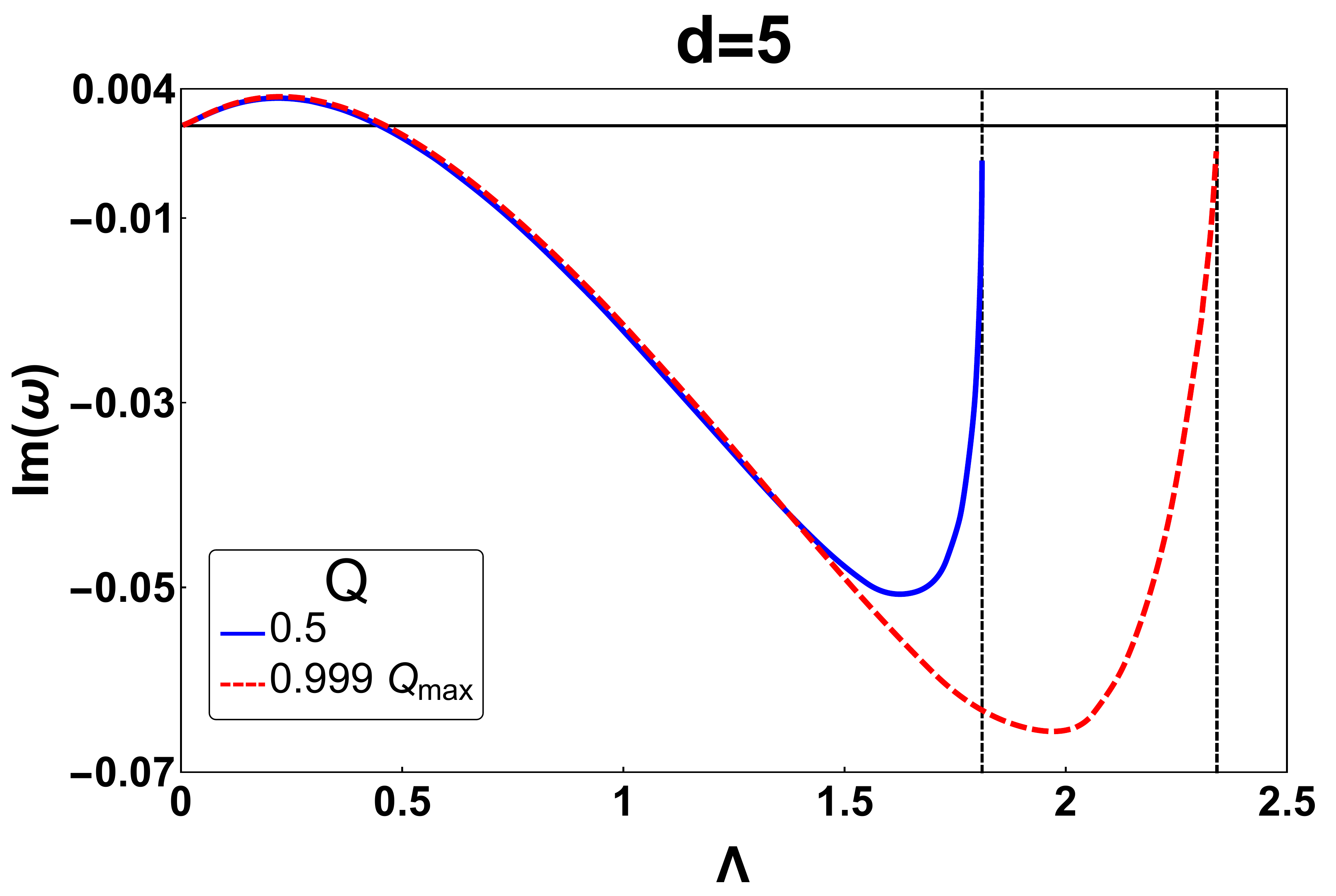}}
\hskip -2ex
\subfigure{\includegraphics[scale=0.17]{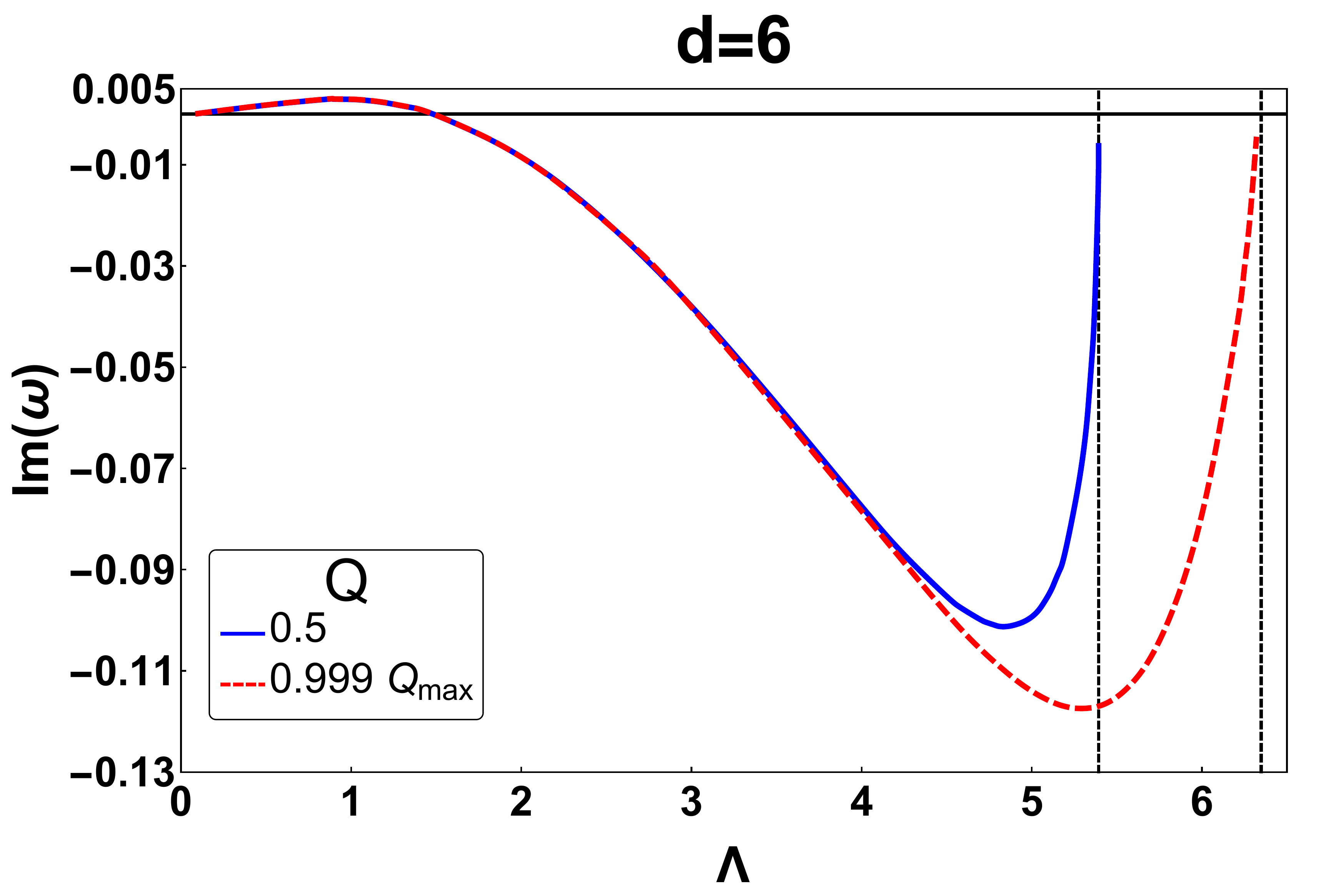}}
\caption{Imaginary parts of $l=0$ charged massless scalar perturbations on a fixed $d-$dimensional RNdS BH with $M=1$ and $qQ=0.4$ (left), $qQ=3$ (middle), $qQ=8$ (right) versus $\Lambda$. The horizontal line designates when $\text{Im}(\omega)=0$ and the vertical dashed lines designate the extremal cosmological constants for each choice of parameters.}
\label{param}
\end{figure*}

Such a peak seems to occur for sufficiently small cosmological constants till the point where $\Lambda$ reaches a critical value beyond which no instability is observed. In Fig. \ref{param} we present the dependence of the unstable $l=0$ modes to the cosmological constant. In all cases, we recognize two turning points: the first one occurs at the peak of instability, where the imaginary part is maximized for the specific choices of parameters, while the second one occurs at the point where the imaginary part is minimized. Beyond the latter, the imaginary part increases again and tends to $0$ from below. At $\Lambda=\Lambda_\text{max}$, $\text{Im}(\omega)$=0, where $\Lambda_\text{max}$ is the extremal cosmological constant of the BH spacetime. This is in accordance with the analytical results in \cite{Hod:2018fet} (at least for $4-$dimensional RNdS BHs).

Interestingly enough, in Table \ref{table1} we show that the $l=0$ scalar perturbations in $d-$dimensional RNdS BHs with arbitrarily small or large $qQ$, fit the superradiant condition \eqref{suprad}, even when the modes are stable. Moreover, we have found no unstable modes which do not satisfy \eqref{suprad}. This demonstrates that superradiance is necessary but not sufficient for the instability to occur.

\begin{figure*}[ht!]
\subfigure{\includegraphics[scale=0.17]{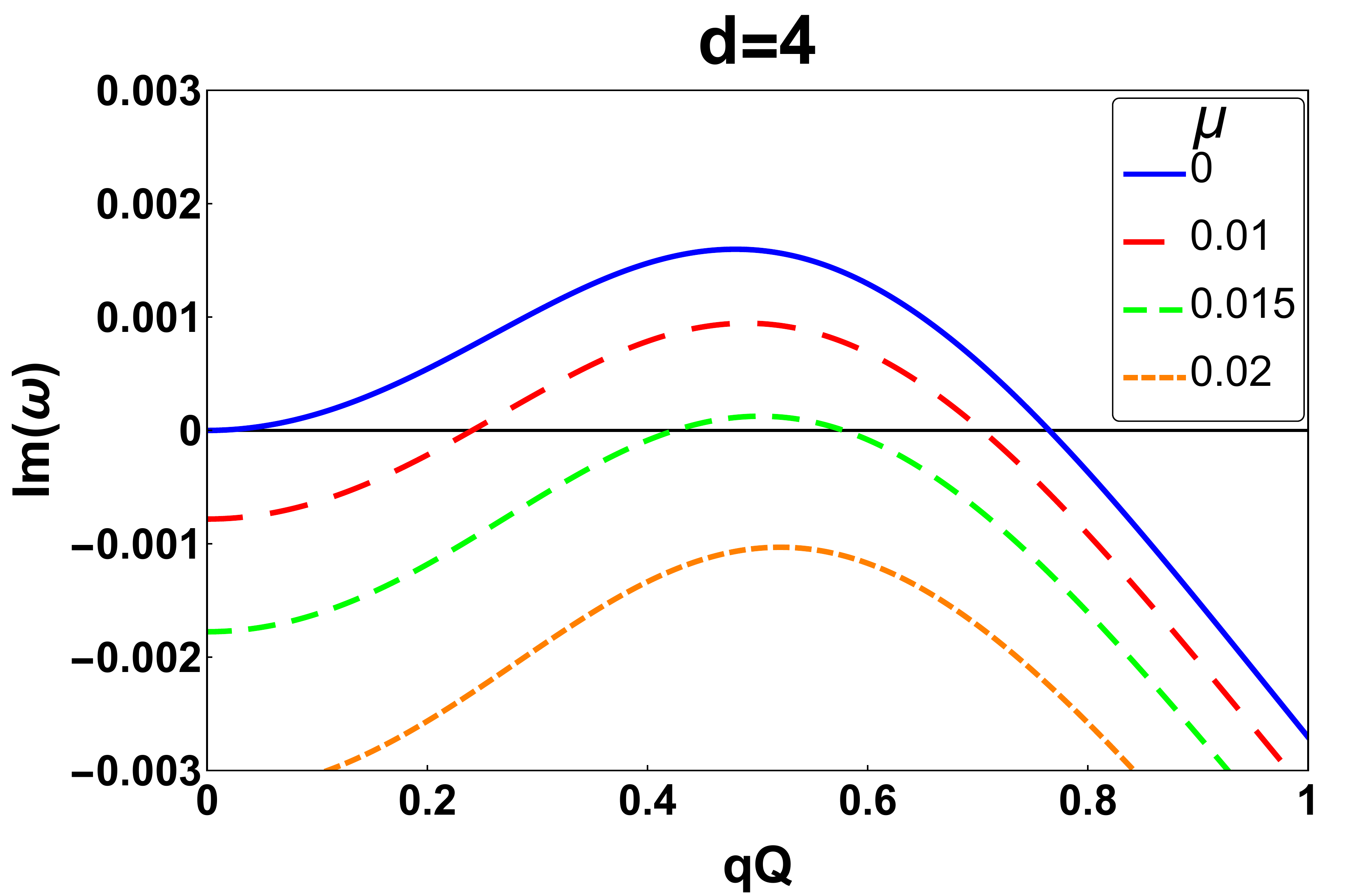}}
\hskip -2ex
\subfigure{\includegraphics[scale=0.17]{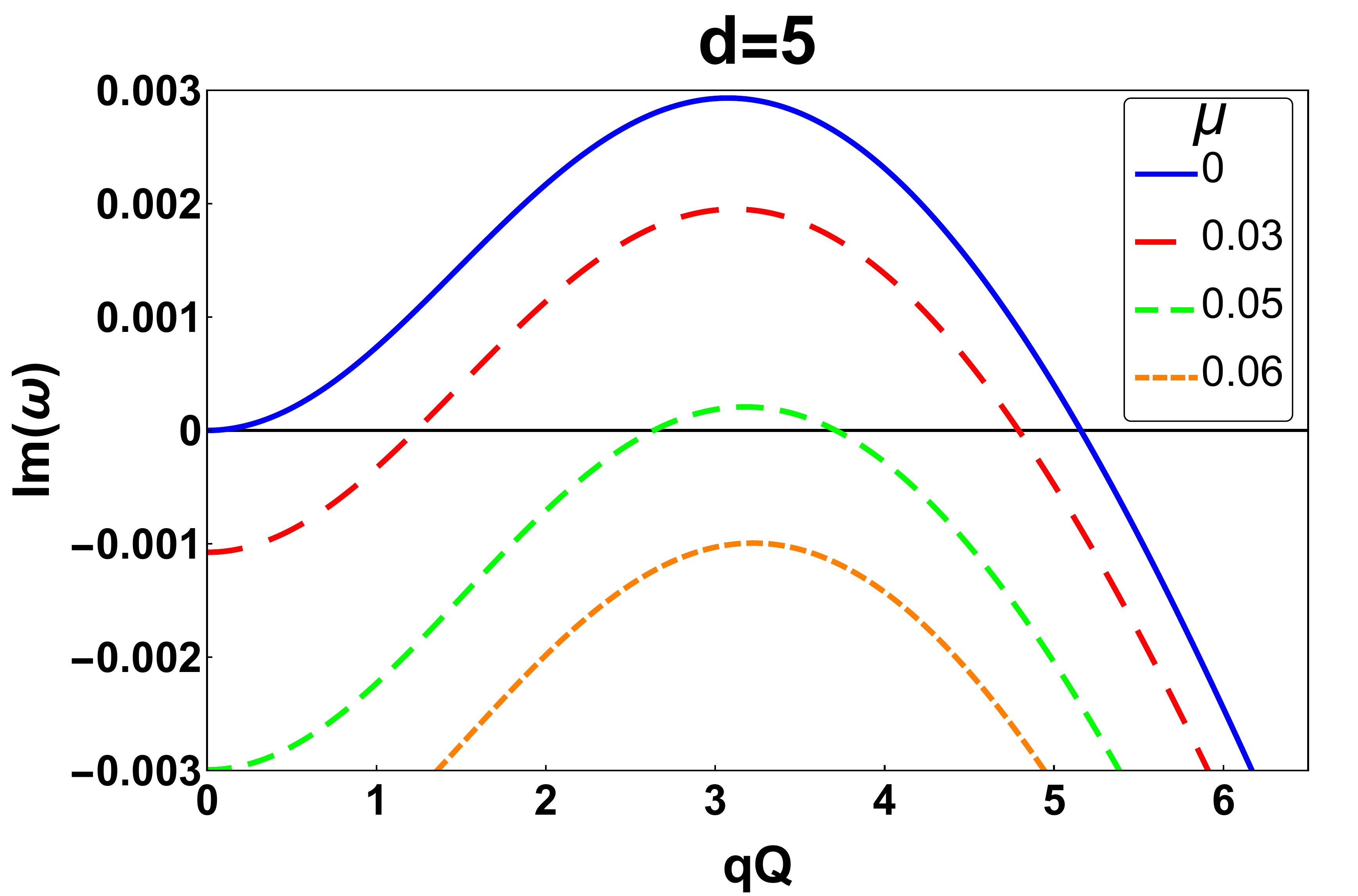}}
\hskip -2ex
\subfigure{\includegraphics[scale=0.17]{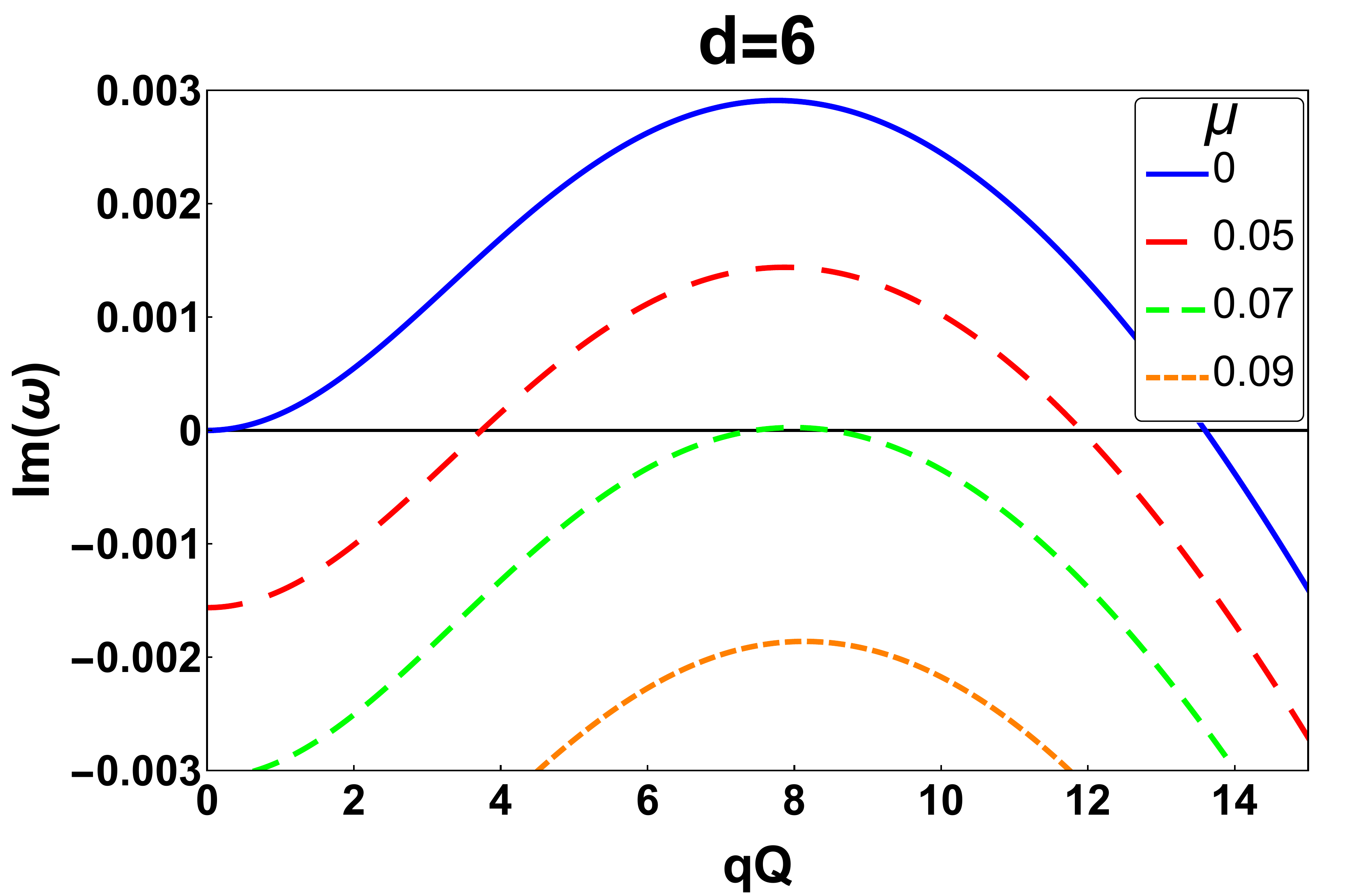}}
\caption{Imaginary parts of $l=0$ charged massive scalar perturbations on a fixed $d-$dimensional RNdS BH with $M=1$, $Q=0.5$ versus the charge coupling $qQ$, for various $\mu$. The cosmological constants used are $\Lambda=0.005$ for $d=4$, $\Lambda=0.25$ for $d=5$ and $\Lambda=1$ for $d=6$.}
\label{massive}
\end{figure*}

%%%%%%%%%%%%%%%%%%%%%%%%%%%%%%%%%%%%%%%%%%%%%%%%%%
\noindent{\bf{\em VI. Unstable massive charged scalar fields.}}
%%%%%%%%%%%%%%%%%%%%%%%%%%%%%%%%%%%%%%%%%%%%%%%%%%
The addition of a non-zero scalar field mass $\mu$ affects the unstable modes in a manner shown in Fig. \ref{massive}. The $\text{Im}(\omega)$ shift downwards with the increment of $\mu$ which leads to increasingly smaller regions of $qQ$ where unstable modes exist. 
%In fact, the "unstable" $l=0$ family will initially originate from stable purely imaginary modes for $qQ=0$. 
After a finite $\mu$, perturbations become stable for all $qQ$. This is due to the upwards shift of (\ref{RNdS_general potential}) as $\mu$ increases which eliminates the potential well formed for $\mu=0$. 

From Fig. \ref{massive} we realize that two critical charge couplings exist, $qQ_c(\text{min})$ beyond which linear instabilities arise and $qQ_c(\text{max})$ beyond which stability is restored. As expected, $qQ_c(\text{min})=0$ and $qQ_c(\text{max})$ is maximized for $\mu=0$. With the increment of $\mu$, $qQ_c(\text{min})$ increases and $qQ_c(\text{max})$ decreases until a finite mass where they coincide. Beyond that point they both vanish and all $l=0$ modes are stable.

Curiously enough, unstable QNMs with non-zero mass exist in various regions of the parameter space and, even more strikingly, they still satisfy (\ref{suprad}). In Table \ref{table2} we show that the addition of a mass term can still lead to unstable and stable superradiant modes. The saturation of the instability with the increment of $\mu$ can be explained by the fact that the gravitational attraction between the massive field and the BH becomes dominant for large enough $\mu$ if one compares it with the electric repulsion between the BH and scalar field charges $qQ$.

\begin{table*}[ht]
\centering
\scalebox{0.6}{
\begin{tabular}{||c| c | c | c ||} 
\hline
  \multicolumn{4}{||c||}{$d=4$} \\
   \hline
\hline
  \multicolumn{4}{||c||}{$\Lambda=0.005$} \\
   \hline
    $qQ$ & $\omega$ & $\Phi(r_c)$ &$\Phi(r_+)$ \\ [0.5ex]
   \hline
    0.005 & 0.00023 +  3.7$\times$$10^{-7}$i  & 0.00021   &0.00266 \\
   \hline
    0.05 & 0.00229 + 0.00004 i  & 0.00213   &0.02663 \\
   \hline
   0.5 & 0.02577 + 0.00159 i  & 0.02134   &0.26625 \\
   \hline
   1 & 0.05322 - 0.00271 i & 0.04268&0.53251 \\ 
   \hline
   5 &0.21907 - 0.01736 i & 0.21338& 2.66253\\
   \hline
   10 &0.42966 - 0.01832 i &0.42677 &5.32507 \\
   \hline\hline
   \multicolumn{4}{||c||}{$\Lambda=0.05$} \\
    \hline
     $qQ$ & $\omega$ & $\Phi(r_c)$ &$\Phi(r_+)$ \\ [0.5ex]
   \hline
    0.005 & 0.00092 - 3.2$\times$$10^{-7}$i  & 0.00077   &0.00249 \\
    \hline
    0.05 & 0.00925 - 0.00003 i  & 0.00773   &0.02486 \\  
    \hline
    0.5 &  0.09222 - 0.00430 i & 0.07734   & 0.24859 \\
    \hline
    1 & 0.17777 - 0.01595 i &0.15468 &0.49719 \\ 
    \hline
    5 &0.78240 - 0.04099 i &0.77338 &2.48594 \\
    \hline
    10 &1.55129 - 0.04204 I &1.54676 &4.97188 \\
    \hline
\end{tabular}
}
\scalebox{0.6}{
\begin{tabular}{||c| c | c | c ||} 
\hline
  \multicolumn{4}{||c||}{$d=5$} \\
   \hline
\hline
  \multicolumn{4}{||c||}{$\Lambda=0.25$} \\
   \hline
    $qQ$ & $\omega$ & $\Phi(r_c)$ &$\Phi(r_+)$ \\ [0.5ex] 
  \hline
   0.05 &0.00081 + 2$\times 10^{-6}$ i &0.00069  &0.01900 \\
   \hline
   0.5  &0.00818 + 0.00020 i  &0.00688  &0.18998 \\
   \hline
   1   &0.01656 + 0.00073 i  & 0.01377 &0.37996 \\
   \hline
   5 & 0.09344 + 0.00040 i &0.06884 & 1.89978\\ 
   \hline
   10 &0.18206 - 0.01759 i &0.13768 &3.79956 \\
   \hline
   15 &0.25790 - 0.03625 i & 0.20651&5.69934 \\
   \hline\hline
   \multicolumn{4}{||c||}{$\Lambda=0.75$} \\
    \hline
     $qQ$ & $\omega$ & $\Phi(r_c)$ &$\Phi(r_+)$ \\ [0.5ex] 
    \hline
       0.05 & 0.00293 - $4\times 10^{-7}$ i &0.002260 &0.01735 \\
       \hline
       0.5 & 0.02945 - 0.00006 i  & 0.02260   &0.17353 \\
       \hline
       1 &  0.05922 - 0.00042 i & 0.04521   &0.34707 \\
       \hline
       5 & 0.28945 - 0.03062 i &0.22604 &1.73533 \\ 
       \hline
       10 &0.52736 - 0.07913 i & 0.45208&3.47066 \\
       \hline
       15 &0.74441 - 0.10774 i &0.67812 &5.20598\\
    \hline
\end{tabular}
}
\scalebox{0.6}{
\begin{tabular}{||c| c | c | c ||} 
\hline
  \multicolumn{4}{||c||}{$d=6$} \\
   \hline
\hline
  \multicolumn{4}{||c||}{$\Lambda=1$} \\
   \hline
    $qQ$ & $\omega$ & $\Phi(r_c)$ &$\Phi(r_+)$ \\ [0.5ex] 
  \hline
   0.05 &0.00032 + 4$\times 10^{-7}$ i &0.00026  &0.01630 \\
   \hline
   0.5  &0.00321 + 0.00004 i  &0.00258  &0.16297 \\
   \hline
   1   &0.00644 + 0.00015 i  & 0.00515 &0.32594 \\
   \hline
   5 & 0.03428 + 0.00222 i &0.02576 & 1.62972\\ 
   \hline
   10 &0.07291 + 0.00245 i &0.05152 &3.25944 \\
   \hline
   20 &0.14819 - 0.00775 i & 0.10304&6.51887 \\
   \hline\hline
   \multicolumn{4}{||c||}{$\Lambda=2$} \\
    \hline
     $qQ$ & $\omega$ & $\Phi(r_c)$ &$\Phi(r_+)$ \\ [0.5ex] 
    \hline
       0.05 & 0.00103 + $5\times 10^{-7}$ i &0.00076 &0.01505 \\
       \hline
       0.5 & 0.01028 + 0.00005 i  & 0.00764   &0.15054 \\
       \hline
       1 &  0.02062 - 0.00018 i & 0.01527   &0.30108 \\
       \hline
       5 & 0.10765 - 0.00084 i &0.07635 &1.50539 \\ 
       \hline
       10 &0.21518 - 0.01597 i & 0.15271&3.01079 \\
       \hline
       20 &0.40031 - 0.06012 i &0.30541 &6.02159\\
    \hline
\end{tabular}
}
\caption{Dominant $l=0$ massless charged scalar modes of $d-$dimensional RNdS spacetime with $M=1$, $Q=0.5$ for various parameters. All modes shown satisfy the superradiant condition (\ref{suprad}).}
\label{table1}
\end{table*}
\begin{table}[ht!]
\centering
\scalebox{0.8}{
\begin{tabular}{||c| c | c | c ||} 
\hline
  \multicolumn{4}{||c||}{$d=4, \Lambda=0.005$, $qQ=0.45$} \\
   \hline
    $\mu $ & $\omega$ & $\Phi(r_c)$ &$\Phi(r_+)$ \\ [0.5ex] 
  \hline
   $10^{-3}$ &0.02305 + 0.00173 i &0.01919  &0.42925 \\
   \hline
   $10^{-2}$  & 0.02280 + 0.00107 i  &0.01919  &0.42925 \\
   \hline
   $3\times 10^{-2}$ &0.02054 - 0.00458 i  &0.01919  &0.42925 \\ 
   \hline
   $5\times 10^{-2}$ &0.01263 - 0.01796 i &0.01919  &0.42925 \\
   \hline
\end{tabular}
}
\scalebox{0.8}{
\begin{tabular}{||c| c | c | c ||} 
\hline
  \multicolumn{4}{||c||}{$d=5, \Lambda=0.25$, $qQ=3$} \\
   \hline
    $\mu $ & $\omega$ & $\Phi(r_c)$ &$\Phi(r_+)$ \\ [0.5ex] 
  \hline
   $10^{-3}$ &0.05381 + 0.00293 i &0.04130  &1.13987 \\
   \hline
   $10^{-2}$  & 0.05379 + 0.00282 i  &0.04130  &1.13987 \\
   \hline
   $ 10^{-1}$ &0.05093 - 0.00819 i  &0.04130  &1.13987 \\ 
   \hline
   $2\times 10^{-1}$ &0.04100 - 0.04427 i &0.04130  &1.13987 \\
   \hline
\end{tabular}
}
\scalebox{0.8}{
\begin{tabular}{||c| c | c | c ||} 
\hline
  \multicolumn{4}{||c||}{$d=6, \Lambda=1$, $qQ=7$} \\
   \hline
    $\mu$ & $\omega$ & $\Phi(r_c)$ &$\Phi(r_+)$ \\ [0.5ex] 
  \hline
   $10^{-3}$ &0.04949 + 0.00286 i &0.03606  &2.28161 \\
   \hline
   $10^{-2}$  & 0.04948 + 0.00280 i  &0.03606  &2.28161 \\
   \hline
   $4\times 10^{-1}$ &0.04496 - 0.02141 i  &0.03606  &2.28161 \\ 
   \hline
   $4\times 10^{-1}$ &0.03005 - 0.10116 i &0.03606  &2.28161 \\
   \hline
\end{tabular}
}
\caption{Dominant $l=0$ massive charged scalar modes of $d-$dimensional RNdS spacetime with $M=1$, $Q=0.5$ for various parameters.}
\label{table2}
\end{table}

%%%%%%%%%%%%%%%%%%%%%%%%%%%%%%%%%%%%%%%%%%%%%%%%%%
\noindent{\bf{\em VII. Conclusions.}}
%%%%%%%%%%%%%%%%%%%%%%%%%%%%%%%%%%%%%%%%%%%%%%%%%%
In the present work, we study a dynamical instability emerging from a spherically-symmetric charged scalar perturbation propagating on a fixed $d-$dimensional RNdS background. Such an instability has a superradiant nature, therefore the scattered wave's amplitude can grow in expense of the BH's electromagnetic energy.

By performing a thorough frequency-domain analysis of higher-dimensional RNdS BHs under charged scalar perturbations we realize that the source of instability is directly linked with the accelerated expansion of the dS Universe, as well as the QNMs of pure $d-$dimensional de Sitter space. 

As the PS vibrates when we probe it, leaking out energy in a particular manner described by QNMs, so the cosmological horizon fluctuates. These new "oscillations" have a distinct nature, comparing to any other QN oscillatory mode. They originate from purely imaginary modes, very well approximated by the pure de Sitter space QNMs and even exhibit a stationary mode $\omega=0$ for any dimension with vanishing angular momentum. 

When charge is introduced to the scalar field, the "zero-mode" evolves to a complex QNM with positive imaginary part and monotonously increasing real part. For a finite region of the charge coupling $qQ$ the family remains unstable which lead to a growing profile of the perturbation with respect to time. Such modes satisfy the superradiant condition even when stable configurations occur. 

With the increment of dimensions, the instability is amplified, leading to larger regions in the parameter space where the unstable modes occur. Interestingly, even though the introduction of mass stabilizes the system, there are still regions in the parameter space where the family remains superradiantly unstable. 

An open, and still interesting, problem is the nonlinear development of such a system to grasp the end-state of the dynamically evolving BH (see \cite{Luna:2018jfk} for the neutral self-gravitating scalar field case).  A huge challenge in such nonlinear evolutions is the very large timescale of the instability which requires highly precise numerical developments. In any case, since the increment of dimensions amplifies the instability, it would be more feasible for such an instability to be tested in higher-dimensional RNdS spacetime non-linearly and realize if it leads to a novel scalarized BH or to the evacuation of all matter.

%%%%%%%%%%%%%%%%%%%%%%%%%%%%%%%%%%%%%%%%%%%%%%%%%%
\noindent{\bf{\em Acknowledgments.}}
%%%%%%%%%%%%%%%%%%%%%%%%%%%%%%%%%%%%%%%%%%%%%%%%%%
The author would like to thank Vitor Cardoso, Aron Jansen, Rodrigo Vicente and Rodrigo Fontana for helpful discussions. The author would especially like to thank Raimon Luna and Miguel Zilh\~{a}o for providing time evolution data that confirmed the validity of the spectral analysis. K.D. acknowledges financial support provided under the European Union's H2020 ERC 
Consolidator Grant ``Matter and strong-field gravity: New frontiers in Einstein's theory'' grant 
agreement no. MaGRaTh--646597. 
K.D. would like to acknowledge networking support by the GWverse COST Action CA16104, ``Black holes, gravitational waves and fundamental physics.''

\bibliography{references}

\begin{thebibliography}{75}%
\makeatletter
\providecommand \@ifxundefined [1]{%
 \@ifx{#1\undefined}
}%
\providecommand \@ifnum [1]{%
 \ifnum #1\expandafter \@firstoftwo
 \else \expandafter \@secondoftwo
 \fi
}%
\providecommand \@ifx [1]{%
 \ifx #1\expandafter \@firstoftwo
 \else \expandafter \@secondoftwo
 \fi
}%
\providecommand \natexlab [1]{#1}%
\providecommand \enquote  [1]{``#1''}%
\providecommand \bibnamefont  [1]{#1}%
\providecommand \bibfnamefont [1]{#1}%
\providecommand \citenamefont [1]{#1}%
\providecommand \href@noop [0]{\@secondoftwo}%
\providecommand \href [0]{\begingroup \@sanitize@url \@href}%
\providecommand \@href[1]{\@@startlink{#1}\@@href}%
\providecommand \@@href[1]{\endgroup#1\@@endlink}%
\providecommand \@sanitize@url [0]{\catcode `\\12\catcode `\$12\catcode
  `\&12\catcode `\#12\catcode `\^12\catcode `\_12\catcode `\%12\relax}%
\providecommand \@@startlink[1]{}%
\providecommand \@@endlink[0]{}%
\providecommand \url  [0]{\begingroup\@sanitize@url \@url }%
\providecommand \@url [1]{\endgroup\@href {#1}{\urlprefix }}%
\providecommand \urlprefix  [0]{URL }%
\providecommand \Eprint [0]{\href }%
\providecommand \doibase [0]{http://dx.doi.org/}%
\providecommand \selectlanguage [0]{\@gobble}%
\providecommand \bibinfo  [0]{\@secondoftwo}%
\providecommand \bibfield  [0]{\@secondoftwo}%
\providecommand \translation [1]{[#1]}%
\providecommand \BibitemOpen [0]{}%
\providecommand \bibitemStop [0]{}%
\providecommand \bibitemNoStop [0]{.\EOS\space}%
\providecommand \EOS [0]{\spacefactor3000\relax}%
\providecommand \BibitemShut  [1]{\csname bibitem#1\endcsname}%
\let\auto@bib@innerbib\@empty
%</preamble>
\bibitem [{\citenamefont {Regge}\ and\ \citenamefont {Wheeler}(1957)}]{Regge}%
  \BibitemOpen
  \bibfield  {author} {\bibinfo {author} {\bibfnamefont {T.}~\bibnamefont
  {Regge}}\ and\ \bibinfo {author} {\bibfnamefont {J.~A.}\ \bibnamefont
  {Wheeler}},\ }\href {\doibase 10.1103/PhysRev.108.1063} {\bibfield  {journal}
  {\bibinfo  {journal} {Phys. Rev.}\ }\textbf {\bibinfo {volume} {108}},\
  \bibinfo {pages} {1063} (\bibinfo {year} {1957})}\BibitemShut {NoStop}%
\bibitem [{\citenamefont {Barack}\ \emph {et~al.}(2018)\citenamefont {Barack}
  \emph {et~al.}}]{Barack:2018yly}%
  \BibitemOpen
  \bibfield  {author} {\bibinfo {author} {\bibfnamefont {L.}~\bibnamefont
  {Barack}} \emph {et~al.},\ }\href@noop {} {\  (\bibinfo {year} {2018})},\
  \Eprint {http://arxiv.org/abs/1806.05195} {arXiv:1806.05195 [gr-qc]}
  \BibitemShut {NoStop}%
%%CITATION = ARXIV:1806.05195;%%
\bibitem [{\citenamefont {Pani}(2013)}]{Pani:2013pma}%
  \BibitemOpen
  \bibfield  {author} {\bibinfo {author} {\bibfnamefont {P.}~\bibnamefont
  {Pani}},\ }\bibfield  {booktitle} {\emph {\bibinfo {booktitle} {{Proceedings,
  Spring School on Numerical Relativity and High Energy Physics (NR/HEP2):
  Lisbon, Portugal, March 11-14, 2013}}},\ }\href {\doibase
  10.1142/S0217751X13400186} {\bibfield  {journal} {\bibinfo  {journal} {Int.
  J. Mod. Phys.}\ }\textbf {\bibinfo {volume} {A28}},\ \bibinfo {pages}
  {1340018} (\bibinfo {year} {2013})},\ \Eprint
  {http://arxiv.org/abs/1305.6759} {arXiv:1305.6759 [gr-qc]} \BibitemShut
  {NoStop}%
%%CITATION = ARXIV:1305.6759;%%
\bibitem [{\citenamefont {Kokkotas}\ and\ \citenamefont
  {Schmidt}(1999)}]{Kokkotas:1999bd}%
  \BibitemOpen
  \bibfield  {author} {\bibinfo {author} {\bibfnamefont {K.~D.}\ \bibnamefont
  {Kokkotas}}\ and\ \bibinfo {author} {\bibfnamefont {B.~G.}\ \bibnamefont
  {Schmidt}},\ }\href {\doibase 10.12942/lrr-1999-2} {\bibfield  {journal}
  {\bibinfo  {journal} {Living Rev. Rel.}\ }\textbf {\bibinfo {volume} {2}},\
  \bibinfo {pages} {2} (\bibinfo {year} {1999})},\ \Eprint
  {http://arxiv.org/abs/gr-qc/9909058} {arXiv:gr-qc/9909058 [gr-qc]}
  \BibitemShut {NoStop}%
%%CITATION = GR-QC/9909058;%%
\bibitem [{\citenamefont {Berti}\ \emph {et~al.}(2009)\citenamefont {Berti},
  \citenamefont {Cardoso},\ and\ \citenamefont {Starinets}}]{Berti:2009kk}%
  \BibitemOpen
  \bibfield  {author} {\bibinfo {author} {\bibfnamefont {E.}~\bibnamefont
  {Berti}}, \bibinfo {author} {\bibfnamefont {V.}~\bibnamefont {Cardoso}}, \
  and\ \bibinfo {author} {\bibfnamefont {A.~O.}\ \bibnamefont {Starinets}},\
  }\href {\doibase 10.1088/0264-9381/26/16/163001} {\bibfield  {journal}
  {\bibinfo  {journal} {Class. Quant. Grav.}\ }\textbf {\bibinfo {volume}
  {26}},\ \bibinfo {pages} {163001} (\bibinfo {year} {2009})},\ \Eprint
  {http://arxiv.org/abs/0905.2975} {arXiv:0905.2975 [gr-qc]} \BibitemShut
  {NoStop}%
%%CITATION = ARXIV:0905.2975;%%
\bibitem [{\citenamefont {Konoplya}\ and\ \citenamefont
  {Zhidenko}(2011)}]{Konoplya:2011qq}%
  \BibitemOpen
  \bibfield  {author} {\bibinfo {author} {\bibfnamefont {R.~A.}\ \bibnamefont
  {Konoplya}}\ and\ \bibinfo {author} {\bibfnamefont {A.}~\bibnamefont
  {Zhidenko}},\ }\href {\doibase 10.1103/RevModPhys.83.793} {\bibfield
  {journal} {\bibinfo  {journal} {Rev. Mod. Phys.}\ }\textbf {\bibinfo {volume}
  {83}},\ \bibinfo {pages} {793} (\bibinfo {year} {2011})},\ \Eprint
  {http://arxiv.org/abs/1102.4014} {arXiv:1102.4014 [gr-qc]} \BibitemShut
  {NoStop}%
%%CITATION = ARXIV:1102.4014;%%
\bibitem [{\citenamefont {Penrose}\ and\ \citenamefont
  {Floyd}(1971)}]{Penrose:1971uk}%
  \BibitemOpen
  \bibfield  {author} {\bibinfo {author} {\bibfnamefont {R.}~\bibnamefont
  {Penrose}}\ and\ \bibinfo {author} {\bibfnamefont {R.~M.}\ \bibnamefont
  {Floyd}},\ }\href@noop {} {\bibfield  {journal} {\bibinfo  {journal}
  {Nature}\ }\textbf {\bibinfo {volume} {229}},\ \bibinfo {pages} {177}
  (\bibinfo {year} {1971})}\BibitemShut {NoStop}%
%%CITATION = NATUA,229,177;%%
\bibitem [{\citenamefont {Bekenstein}(1973)}]{Bekenstein:1973mi}%
  \BibitemOpen
  \bibfield  {author} {\bibinfo {author} {\bibfnamefont {J.~D.}\ \bibnamefont
  {Bekenstein}},\ }\href {\doibase 10.1103/PhysRevD.7.949} {\bibfield
  {journal} {\bibinfo  {journal} {Phys. Rev.}\ }\textbf {\bibinfo {volume}
  {D7}},\ \bibinfo {pages} {949} (\bibinfo {year} {1973})}\BibitemShut
  {NoStop}%
%%CITATION = PHRVA,D7,949;%%
\bibitem [{\citenamefont {Brito}\ \emph {et~al.}(2015)\citenamefont {Brito},
  \citenamefont {Cardoso},\ and\ \citenamefont {Pani}}]{Brito:2015oca}%
  \BibitemOpen
  \bibfield  {author} {\bibinfo {author} {\bibfnamefont {R.}~\bibnamefont
  {Brito}}, \bibinfo {author} {\bibfnamefont {V.}~\bibnamefont {Cardoso}}, \
  and\ \bibinfo {author} {\bibfnamefont {P.}~\bibnamefont {Pani}},\ }\href
  {\doibase 10.1007/978-3-319-19000-6} {\bibfield  {journal} {\bibinfo
  {journal} {Lect. Notes Phys.}\ }\textbf {\bibinfo {volume} {906}},\ \bibinfo
  {pages} {pp.1} (\bibinfo {year} {2015})},\ \Eprint
  {http://arxiv.org/abs/1501.06570} {arXiv:1501.06570 [gr-qc]} \BibitemShut
  {NoStop}%
%%CITATION = ARXIV:1501.06570;%%
\bibitem [{\citenamefont {Detweiler}(1980)}]{Detweiler}%
  \BibitemOpen
  \bibfield  {author} {\bibinfo {author} {\bibfnamefont {S.}~\bibnamefont
  {Detweiler}},\ }\href {\doibase 10.1103/PhysRevD.22.2323} {\bibfield
  {journal} {\bibinfo  {journal} {Phys. Rev. D}\ }\textbf {\bibinfo {volume}
  {22}},\ \bibinfo {pages} {2323} (\bibinfo {year} {1980})}\BibitemShut
  {NoStop}%
\bibitem [{\citenamefont {Cardoso}\ and\ \citenamefont
  {Yoshida}(2005)}]{Vitor1}%
  \BibitemOpen
  \bibfield  {author} {\bibinfo {author} {\bibfnamefont {V.}~\bibnamefont
  {Cardoso}}\ and\ \bibinfo {author} {\bibfnamefont {S.}~\bibnamefont
  {Yoshida}},\ }\href {\doibase 10.1088/1126-6708/2005/07/009} {\bibfield
  {journal} {\bibinfo  {journal} {JHEP}\ }\textbf {\bibinfo {volume} {07}},\
  \bibinfo {pages} {009} (\bibinfo {year} {2005})},\ \Eprint
  {http://arxiv.org/abs/hep-th/0502206} {arXiv:hep-th/0502206 [hep-th]}
  \BibitemShut {NoStop}%
%%CITATION = HEP-TH/0502206;%%
\bibitem [{\citenamefont {Dolan}(2007)}]{Vitor2}%
  \BibitemOpen
  \bibfield  {author} {\bibinfo {author} {\bibfnamefont {S.~R.}\ \bibnamefont
  {Dolan}},\ }\href {\doibase 10.1103/PhysRevD.76.084001} {\bibfield  {journal}
  {\bibinfo  {journal} {Phys. Rev.}\ }\textbf {\bibinfo {volume} {D76}},\
  \bibinfo {pages} {084001} (\bibinfo {year} {2007})},\ \Eprint
  {http://arxiv.org/abs/0705.2880} {arXiv:0705.2880 [gr-qc]} \BibitemShut
  {NoStop}%
%%CITATION = ARXIV:0705.2880;%%
\bibitem [{\citenamefont {Cardoso}\ \emph {et~al.}(2004)\citenamefont
  {Cardoso}, \citenamefont {Dias}, \citenamefont {Lemos},\ and\ \citenamefont
  {Yoshida}}]{Vitor3}%
  \BibitemOpen
  \bibfield  {author} {\bibinfo {author} {\bibfnamefont {V.}~\bibnamefont
  {Cardoso}}, \bibinfo {author} {\bibfnamefont {O.~J.~C.}\ \bibnamefont
  {Dias}}, \bibinfo {author} {\bibfnamefont {J.~P.~S.}\ \bibnamefont {Lemos}},
  \ and\ \bibinfo {author} {\bibfnamefont {S.}~\bibnamefont {Yoshida}},\ }\href
  {\doibase 10.1103/PhysRevD.70.049903, 10.1103/PhysRevD.70.044039} {\bibfield
  {journal} {\bibinfo  {journal} {Phys. Rev.}\ }\textbf {\bibinfo {volume}
  {D70}},\ \bibinfo {pages} {044039} (\bibinfo {year} {2004})},\ \bibinfo
  {note} {[Erratum: Phys. Rev.D70,049903(2004)]},\ \Eprint
  {http://arxiv.org/abs/hep-th/0404096} {arXiv:hep-th/0404096 [hep-th]}
  \BibitemShut {NoStop}%
%%CITATION = HEP-TH/0404096;%%
\bibitem [{\citenamefont {Witek}\ \emph {et~al.}(2013)\citenamefont {Witek},
  \citenamefont {Cardoso}, \citenamefont {Ishibashi},\ and\ \citenamefont
  {Sperhake}}]{Vitor4}%
  \BibitemOpen
  \bibfield  {author} {\bibinfo {author} {\bibfnamefont {H.}~\bibnamefont
  {Witek}}, \bibinfo {author} {\bibfnamefont {V.}~\bibnamefont {Cardoso}},
  \bibinfo {author} {\bibfnamefont {A.}~\bibnamefont {Ishibashi}}, \ and\
  \bibinfo {author} {\bibfnamefont {U.}~\bibnamefont {Sperhake}},\ }\href
  {\doibase 10.1103/PhysRevD.87.043513} {\bibfield  {journal} {\bibinfo
  {journal} {Phys. Rev.}\ }\textbf {\bibinfo {volume} {D87}},\ \bibinfo {pages}
  {043513} (\bibinfo {year} {2013})},\ \Eprint {http://arxiv.org/abs/1212.0551}
  {arXiv:1212.0551 [gr-qc]} \BibitemShut {NoStop}%
%%CITATION = ARXIV:1212.0551;%%
\bibitem [{\citenamefont {Brito}\ \emph {et~al.}(2013)\citenamefont {Brito},
  \citenamefont {Cardoso},\ and\ \citenamefont {Pani}}]{Vitor5}%
  \BibitemOpen
  \bibfield  {author} {\bibinfo {author} {\bibfnamefont {R.}~\bibnamefont
  {Brito}}, \bibinfo {author} {\bibfnamefont {V.}~\bibnamefont {Cardoso}}, \
  and\ \bibinfo {author} {\bibfnamefont {P.}~\bibnamefont {Pani}},\ }\href
  {\doibase 10.1103/PhysRevD.88.023514} {\bibfield  {journal} {\bibinfo
  {journal} {Phys. Rev.}\ }\textbf {\bibinfo {volume} {D88}},\ \bibinfo {pages}
  {023514} (\bibinfo {year} {2013})},\ \Eprint {http://arxiv.org/abs/1304.6725}
  {arXiv:1304.6725 [gr-qc]} \BibitemShut {NoStop}%
%%CITATION = ARXIV:1304.6725;%%
\bibitem [{\citenamefont {Ishibashi}\ \emph {et~al.}(2015)\citenamefont
  {Ishibashi}, \citenamefont {Pani}, \citenamefont {Gualtieri},\ and\
  \citenamefont {Cardoso}}]{Kerr1}%
  \BibitemOpen
  \bibfield  {author} {\bibinfo {author} {\bibfnamefont {A.}~\bibnamefont
  {Ishibashi}}, \bibinfo {author} {\bibfnamefont {P.}~\bibnamefont {Pani}},
  \bibinfo {author} {\bibfnamefont {L.}~\bibnamefont {Gualtieri}}, \ and\
  \bibinfo {author} {\bibfnamefont {V.}~\bibnamefont {Cardoso}},\ }\href
  {\doibase 10.1007/JHEP09(2015)209} {\bibfield  {journal} {\bibinfo  {journal}
  {JHEP}\ }\textbf {\bibinfo {volume} {09}},\ \bibinfo {pages} {209} (\bibinfo
  {year} {2015})},\ \Eprint {http://arxiv.org/abs/1507.07079} {arXiv:1507.07079
  [hep-th]} \BibitemShut {NoStop}%
%%CITATION = ARXIV:1507.07079;%%
\bibitem [{\citenamefont {Wang}\ and\ \citenamefont {Herdeiro}(2016)}]{Kerr2}%
  \BibitemOpen
  \bibfield  {author} {\bibinfo {author} {\bibfnamefont {M.}~\bibnamefont
  {Wang}}\ and\ \bibinfo {author} {\bibfnamefont {C.}~\bibnamefont
  {Herdeiro}},\ }\href {\doibase 10.1103/PhysRevD.93.064066} {\bibfield
  {journal} {\bibinfo  {journal} {Phys. Rev.}\ }\textbf {\bibinfo {volume}
  {D93}},\ \bibinfo {pages} {064066} (\bibinfo {year} {2016})},\ \Eprint
  {http://arxiv.org/abs/1512.02262} {arXiv:1512.02262 [gr-qc]} \BibitemShut
  {NoStop}%
%%CITATION = ARXIV:1512.02262;%%
\bibitem [{\citenamefont {East}\ and\ \citenamefont {Pretorius}(2017)}]{Kerr3}%
  \BibitemOpen
  \bibfield  {author} {\bibinfo {author} {\bibfnamefont {W.~E.}\ \bibnamefont
  {East}}\ and\ \bibinfo {author} {\bibfnamefont {F.}~\bibnamefont
  {Pretorius}},\ }\href {\doibase 10.1103/PhysRevLett.119.041101} {\bibfield
  {journal} {\bibinfo  {journal} {Phys. Rev. Lett.}\ }\textbf {\bibinfo
  {volume} {119}},\ \bibinfo {pages} {041101} (\bibinfo {year} {2017})},\
  \Eprint {http://arxiv.org/abs/1704.04791} {arXiv:1704.04791 [gr-qc]}
  \BibitemShut {NoStop}%
%%CITATION = ARXIV:1704.04791;%%
\bibitem [{\citenamefont {Ferreira}\ and\ \citenamefont
  {Herdeiro}(2018)}]{Kerr4}%
  \BibitemOpen
  \bibfield  {author} {\bibinfo {author} {\bibfnamefont {H.~R.~C.}\
  \bibnamefont {Ferreira}}\ and\ \bibinfo {author} {\bibfnamefont {C.~A.~R.}\
  \bibnamefont {Herdeiro}},\ }\href {\doibase 10.1103/PhysRevD.97.084003}
  {\bibfield  {journal} {\bibinfo  {journal} {Phys. Rev.}\ }\textbf {\bibinfo
  {volume} {D97}},\ \bibinfo {pages} {084003} (\bibinfo {year} {2018})},\
  \Eprint {http://arxiv.org/abs/1712.03398} {arXiv:1712.03398 [gr-qc]}
  \BibitemShut {NoStop}%
%%CITATION = ARXIV:1712.03398;%%
\bibitem [{\citenamefont {Degollado}\ \emph {et~al.}(2018)\citenamefont
  {Degollado}, \citenamefont {Herdeiro},\ and\ \citenamefont {Radu}}]{Kerr5}%
  \BibitemOpen
  \bibfield  {author} {\bibinfo {author} {\bibfnamefont {J.~C.}\ \bibnamefont
  {Degollado}}, \bibinfo {author} {\bibfnamefont {C.~A.~R.}\ \bibnamefont
  {Herdeiro}}, \ and\ \bibinfo {author} {\bibfnamefont {E.}~\bibnamefont
  {Radu}},\ }\href {\doibase 10.1016/j.physletb.2018.04.052} {\bibfield
  {journal} {\bibinfo  {journal} {Phys. Lett.}\ }\textbf {\bibinfo {volume}
  {B781}},\ \bibinfo {pages} {651} (\bibinfo {year} {2018})},\ \Eprint
  {http://arxiv.org/abs/1802.07266} {arXiv:1802.07266 [gr-qc]} \BibitemShut
  {NoStop}%
%%CITATION = ARXIV:1802.07266;%%
\bibitem [{\citenamefont {Kolyvaris}\ \emph {et~al.}(2018)\citenamefont
  {Kolyvaris}, \citenamefont {Koukouvaou}, \citenamefont {Machattou},\ and\
  \citenamefont {Papantonopoulos}}]{RN1}%
  \BibitemOpen
  \bibfield  {author} {\bibinfo {author} {\bibfnamefont {T.}~\bibnamefont
  {Kolyvaris}}, \bibinfo {author} {\bibfnamefont {M.}~\bibnamefont
  {Koukouvaou}}, \bibinfo {author} {\bibfnamefont {A.}~\bibnamefont
  {Machattou}}, \ and\ \bibinfo {author} {\bibfnamefont {E.}~\bibnamefont
  {Papantonopoulos}},\ }\href {\doibase 10.1103/PhysRevD.98.024045} {\bibfield
  {journal} {\bibinfo  {journal} {Phys. Rev.}\ }\textbf {\bibinfo {volume}
  {D98}},\ \bibinfo {pages} {024045} (\bibinfo {year} {2018})},\ \Eprint
  {http://arxiv.org/abs/1806.11110} {arXiv:1806.11110 [gr-qc]} \BibitemShut
  {NoStop}%
%%CITATION = ARXIV:1806.11110;%%
\bibitem [{\citenamefont {Menza}\ and\ \citenamefont {Nicolas}(2015)}]{RN2}%
  \BibitemOpen
  \bibfield  {author} {\bibinfo {author} {\bibfnamefont {L.~D.}\ \bibnamefont
  {Menza}}\ and\ \bibinfo {author} {\bibfnamefont {J.-P.}\ \bibnamefont
  {Nicolas}},\ }\href {\doibase 10.1088/0264-9381/32/14/145013} {\bibfield
  {journal} {\bibinfo  {journal} {Class. Quant. Grav.}\ }\textbf {\bibinfo
  {volume} {32}},\ \bibinfo {pages} {145013} (\bibinfo {year} {2015})},\
  \Eprint {http://arxiv.org/abs/1411.3988} {arXiv:1411.3988 [math-ph]}
  \BibitemShut {NoStop}%
%%CITATION = ARXIV:1411.3988;%%
\bibitem [{\citenamefont {Benone}\ and\ \citenamefont {Crispino}(2016)}]{RN3}%
  \BibitemOpen
  \bibfield  {author} {\bibinfo {author} {\bibfnamefont {C.~L.}\ \bibnamefont
  {Benone}}\ and\ \bibinfo {author} {\bibfnamefont {L.~C.~B.}\ \bibnamefont
  {Crispino}},\ }\href {\doibase 10.1103/PhysRevD.93.024028} {\bibfield
  {journal} {\bibinfo  {journal} {Phys. Rev.}\ }\textbf {\bibinfo {volume}
  {D93}},\ \bibinfo {pages} {024028} (\bibinfo {year} {2016})},\ \Eprint
  {http://arxiv.org/abs/1511.02634} {arXiv:1511.02634 [gr-qc]} \BibitemShut
  {NoStop}%
%%CITATION = ARXIV:1511.02634;%%
\bibitem [{\citenamefont {Huang}\ \emph {et~al.}(2017)\citenamefont {Huang},
  \citenamefont {Liu},\ and\ \citenamefont {Li}}]{RN4}%
  \BibitemOpen
  \bibfield  {author} {\bibinfo {author} {\bibfnamefont {Y.}~\bibnamefont
  {Huang}}, \bibinfo {author} {\bibfnamefont {D.-J.}\ \bibnamefont {Liu}}, \
  and\ \bibinfo {author} {\bibfnamefont {X.-Z.}\ \bibnamefont {Li}},\ }\href
  {\doibase 10.1142/S0218271817501413} {\bibfield  {journal} {\bibinfo
  {journal} {Int. J. Mod. Phys.}\ }\textbf {\bibinfo {volume} {D26}},\ \bibinfo
  {pages} {1750141} (\bibinfo {year} {2017})},\ \Eprint
  {http://arxiv.org/abs/1606.00100} {arXiv:1606.00100 [gr-qc]} \BibitemShut
  {NoStop}%
%%CITATION = ARXIV:1606.00100;%%
\bibitem [{\citenamefont {Gonz\'{a}lez}\ \emph {et~al.}(2017)\citenamefont
  {Gonz\'{a}lez}, \citenamefont {Papantonopoulos}, \citenamefont {Saavedra},\
  and\ \citenamefont {V\'{a}squez}}]{RN5}%
  \BibitemOpen
  \bibfield  {author} {\bibinfo {author} {\bibfnamefont {P.~A.}\ \bibnamefont
  {Gonz\'{a}lez}}, \bibinfo {author} {\bibfnamefont {E.}~\bibnamefont
  {Papantonopoulos}}, \bibinfo {author} {\bibfnamefont {J.}~\bibnamefont
  {Saavedra}}, \ and\ \bibinfo {author} {\bibfnamefont {Y.}~\bibnamefont
  {V\'{a}squez}},\ }\href {\doibase 10.1103/PhysRevD.95.064046} {\bibfield
  {journal} {\bibinfo  {journal} {Phys. Rev.}\ }\textbf {\bibinfo {volume}
  {D95}},\ \bibinfo {pages} {064046} (\bibinfo {year} {2017})},\ \Eprint
  {http://arxiv.org/abs/1702.00439} {arXiv:1702.00439 [gr-qc]} \BibitemShut
  {NoStop}%
%%CITATION = ARXIV:1702.00439;%%
\bibitem [{\citenamefont {Kolyvaris}\ and\ \citenamefont
  {Papantonopoulos}(2017)}]{RN6}%
  \BibitemOpen
  \bibfield  {author} {\bibinfo {author} {\bibfnamefont {T.}~\bibnamefont
  {Kolyvaris}}\ and\ \bibinfo {author} {\bibfnamefont {E.}~\bibnamefont
  {Papantonopoulos}},\ }\href@noop {} {\  (\bibinfo {year} {2017})},\ \Eprint
  {http://arxiv.org/abs/1702.04618} {arXiv:1702.04618 [gr-qc]} \BibitemShut
  {NoStop}%
%%CITATION = ARXIV:1702.04618;%%
\bibitem [{\citenamefont {Huang}\ \emph {et~al.}(2018)\citenamefont {Huang},
  \citenamefont {Liu}, \citenamefont {Zhai},\ and\ \citenamefont {Li}}]{KN1}%
  \BibitemOpen
  \bibfield  {author} {\bibinfo {author} {\bibfnamefont {Y.}~\bibnamefont
  {Huang}}, \bibinfo {author} {\bibfnamefont {D.-J.}\ \bibnamefont {Liu}},
  \bibinfo {author} {\bibfnamefont {X.-h.}\ \bibnamefont {Zhai}}, \ and\
  \bibinfo {author} {\bibfnamefont {X.-z.}\ \bibnamefont {Li}},\ }\href
  {\doibase 10.1103/PhysRevD.98.025021} {\bibfield  {journal} {\bibinfo
  {journal} {Phys. Rev.}\ }\textbf {\bibinfo {volume} {D98}},\ \bibinfo {pages}
  {025021} (\bibinfo {year} {2018})},\ \Eprint
  {http://arxiv.org/abs/1807.06263} {arXiv:1807.06263 [gr-qc]} \BibitemShut
  {NoStop}%
%%CITATION = ARXIV:1807.06263;%%
\bibitem [{\citenamefont {Benone}\ and\ \citenamefont {Crispino}(2019)}]{KN2}%
  \BibitemOpen
  \bibfield  {author} {\bibinfo {author} {\bibfnamefont {C.~L.}\ \bibnamefont
  {Benone}}\ and\ \bibinfo {author} {\bibfnamefont {L.~C.~B.}\ \bibnamefont
  {Crispino}},\ }\href {\doibase 10.1103/PhysRevD.99.044009} {\bibfield
  {journal} {\bibinfo  {journal} {Phys. Rev.}\ }\textbf {\bibinfo {volume}
  {D99}},\ \bibinfo {pages} {044009} (\bibinfo {year} {2019})},\ \Eprint
  {http://arxiv.org/abs/1901.05592} {arXiv:1901.05592 [gr-qc]} \BibitemShut
  {NoStop}%
%%CITATION = ARXIV:1901.05592;%%
\bibitem [{\citenamefont {Vicente}\ \emph {et~al.}(2018)\citenamefont
  {Vicente}, \citenamefont {Cardoso},\ and\ \citenamefont
  {Lopes}}]{Vicente:2018mxl}%
  \BibitemOpen
  \bibfield  {author} {\bibinfo {author} {\bibfnamefont {R.}~\bibnamefont
  {Vicente}}, \bibinfo {author} {\bibfnamefont {V.}~\bibnamefont {Cardoso}}, \
  and\ \bibinfo {author} {\bibfnamefont {J.~C.}\ \bibnamefont {Lopes}},\ }\href
  {\doibase 10.1103/PhysRevD.97.084032} {\bibfield  {journal} {\bibinfo
  {journal} {Phys. Rev.}\ }\textbf {\bibinfo {volume} {D97}},\ \bibinfo {pages}
  {084032} (\bibinfo {year} {2018})},\ \Eprint
  {http://arxiv.org/abs/1803.08060} {arXiv:1803.08060 [gr-qc]} \BibitemShut
  {NoStop}%
%%CITATION = ARXIV:1803.08060;%%
\bibitem [{\citenamefont {Cardoso}\ \emph {et~al.}(2015)\citenamefont
  {Cardoso}, \citenamefont {Brito},\ and\ \citenamefont {Rosa}}]{stars1}%
  \BibitemOpen
  \bibfield  {author} {\bibinfo {author} {\bibfnamefont {V.}~\bibnamefont
  {Cardoso}}, \bibinfo {author} {\bibfnamefont {R.}~\bibnamefont {Brito}}, \
  and\ \bibinfo {author} {\bibfnamefont {J.~L.}\ \bibnamefont {Rosa}},\ }\href
  {\doibase 10.1103/PhysRevD.91.124026} {\bibfield  {journal} {\bibinfo
  {journal} {Phys. Rev.}\ }\textbf {\bibinfo {volume} {D91}},\ \bibinfo {pages}
  {124026} (\bibinfo {year} {2015})},\ \Eprint
  {http://arxiv.org/abs/1505.05509} {arXiv:1505.05509 [gr-qc]} \BibitemShut
  {NoStop}%
%%CITATION = ARXIV:1505.05509;%%
\bibitem [{\citenamefont {Cardoso}\ \emph {et~al.}(2017)\citenamefont
  {Cardoso}, \citenamefont {Pani},\ and\ \citenamefont {Yu}}]{stars2}%
  \BibitemOpen
  \bibfield  {author} {\bibinfo {author} {\bibfnamefont {V.}~\bibnamefont
  {Cardoso}}, \bibinfo {author} {\bibfnamefont {P.}~\bibnamefont {Pani}}, \
  and\ \bibinfo {author} {\bibfnamefont {T.-T.}\ \bibnamefont {Yu}},\ }\href
  {\doibase 10.1103/PhysRevD.95.124056} {\bibfield  {journal} {\bibinfo
  {journal} {Phys. Rev.}\ }\textbf {\bibinfo {volume} {D95}},\ \bibinfo {pages}
  {124056} (\bibinfo {year} {2017})},\ \Eprint
  {http://arxiv.org/abs/1704.06151} {arXiv:1704.06151 [gr-qc]} \BibitemShut
  {NoStop}%
%%CITATION = ARXIV:1704.06151;%%
\bibitem [{\citenamefont {Maggio}\ \emph {et~al.}(2019)\citenamefont {Maggio},
  \citenamefont {Cardoso}, \citenamefont {Dolan},\ and\ \citenamefont
  {Pani}}]{Maggio:2018ivz}%
  \BibitemOpen
  \bibfield  {author} {\bibinfo {author} {\bibfnamefont {E.}~\bibnamefont
  {Maggio}}, \bibinfo {author} {\bibfnamefont {V.}~\bibnamefont {Cardoso}},
  \bibinfo {author} {\bibfnamefont {S.~R.}\ \bibnamefont {Dolan}}, \ and\
  \bibinfo {author} {\bibfnamefont {P.}~\bibnamefont {Pani}},\ }\href {\doibase
  10.1103/PhysRevD.99.064007} {\bibfield  {journal} {\bibinfo  {journal} {Phys.
  Rev.}\ }\textbf {\bibinfo {volume} {D99}},\ \bibinfo {pages} {064007}
  (\bibinfo {year} {2019})},\ \Eprint {http://arxiv.org/abs/1807.08840}
  {arXiv:1807.08840 [gr-qc]} \BibitemShut {NoStop}%
%%CITATION = ARXIV:1807.08840;%%
\bibitem [{\citenamefont {Chandrasekhar}\ and\ \citenamefont
  {Detweiler}(1975)}]{Chandrasekhar:1975zza}%
  \BibitemOpen
  \bibfield  {author} {\bibinfo {author} {\bibfnamefont {S.}~\bibnamefont
  {Chandrasekhar}}\ and\ \bibinfo {author} {\bibfnamefont {S.~L.}\ \bibnamefont
  {Detweiler}},\ }\href {\doibase 10.1098/rspa.1975.0112} {\bibfield  {journal}
  {\bibinfo  {journal} {Proc. Roy. Soc. Lond.}\ }\textbf {\bibinfo {volume}
  {A344}},\ \bibinfo {pages} {441} (\bibinfo {year} {1975})}\BibitemShut
  {NoStop}%
%%CITATION = PRSLA,A344,441;%%
\bibitem [{\citenamefont {Ferrari}\ and\ \citenamefont
  {Mashhoon}(1984)}]{PhysRevLett.52.1361}%
  \BibitemOpen
  \bibfield  {author} {\bibinfo {author} {\bibfnamefont {V.}~\bibnamefont
  {Ferrari}}\ and\ \bibinfo {author} {\bibfnamefont {B.}~\bibnamefont
  {Mashhoon}},\ }\href {\doibase 10.1103/PhysRevLett.52.1361} {\bibfield
  {journal} {\bibinfo  {journal} {Phys. Rev. Lett.}\ }\textbf {\bibinfo
  {volume} {52}},\ \bibinfo {pages} {1361} (\bibinfo {year}
  {1984})}\BibitemShut {NoStop}%
\bibitem [{\citenamefont {Furuhashi}\ and\ \citenamefont
  {Nambu}(2004)}]{Furuhashi:2004jk}%
  \BibitemOpen
  \bibfield  {author} {\bibinfo {author} {\bibfnamefont {H.}~\bibnamefont
  {Furuhashi}}\ and\ \bibinfo {author} {\bibfnamefont {Y.}~\bibnamefont
  {Nambu}},\ }\href {\doibase 10.1143/PTP.112.983} {\bibfield  {journal}
  {\bibinfo  {journal} {Prog. Theor. Phys.}\ }\textbf {\bibinfo {volume}
  {112}},\ \bibinfo {pages} {983} (\bibinfo {year} {2004})},\ \Eprint
  {http://arxiv.org/abs/gr-qc/0402037} {arXiv:gr-qc/0402037 [gr-qc]}
  \BibitemShut {NoStop}%
%%CITATION = GR-QC/0402037;%%
\bibitem [{\citenamefont {Konoplya}\ and\ \citenamefont
  {Zhidenko}(2009)}]{Konoplya:2008au}%
  \BibitemOpen
  \bibfield  {author} {\bibinfo {author} {\bibfnamefont {R.~A.}\ \bibnamefont
  {Konoplya}}\ and\ \bibinfo {author} {\bibfnamefont {A.}~\bibnamefont
  {Zhidenko}},\ }\href {\doibase 10.1103/PhysRevLett.103.161101} {\bibfield
  {journal} {\bibinfo  {journal} {Phys. Rev. Lett.}\ }\textbf {\bibinfo
  {volume} {103}},\ \bibinfo {pages} {161101} (\bibinfo {year} {2009})},\
  \Eprint {http://arxiv.org/abs/0809.2822} {arXiv:0809.2822 [hep-th]}
  \BibitemShut {NoStop}%
%%CITATION = ARXIV:0809.2822;%%
\bibitem [{\citenamefont {Dolan}(2013)}]{Dolan:2012yt}%
  \BibitemOpen
  \bibfield  {author} {\bibinfo {author} {\bibfnamefont {S.~R.}\ \bibnamefont
  {Dolan}},\ }\href {\doibase 10.1103/PhysRevD.87.124026} {\bibfield  {journal}
  {\bibinfo  {journal} {Phys. Rev.}\ }\textbf {\bibinfo {volume} {D87}},\
  \bibinfo {pages} {124026} (\bibinfo {year} {2013})},\ \Eprint
  {http://arxiv.org/abs/1212.1477} {arXiv:1212.1477 [gr-qc]} \BibitemShut
  {NoStop}%
%%CITATION = ARXIV:1212.1477;%%
\bibitem [{\citenamefont {Destounis}\ \emph {et~al.}(2018)\citenamefont
  {Destounis}, \citenamefont {Panotopoulos},\ and\ \citenamefont
  {Rinc\'{o}n}}]{Destounis:2018utr}%
  \BibitemOpen
  \bibfield  {author} {\bibinfo {author} {\bibfnamefont {K.}~\bibnamefont
  {Destounis}}, \bibinfo {author} {\bibfnamefont {G.}~\bibnamefont
  {Panotopoulos}}, \ and\ \bibinfo {author} {\bibfnamefont {n.}~\bibnamefont
  {Rinc\'{o}n}},\ }\href {\doibase 10.1140/epjc/s10052-018-5576-8} {\bibfield
  {journal} {\bibinfo  {journal} {Eur. Phys. J.}\ }\textbf {\bibinfo {volume}
  {C78}},\ \bibinfo {pages} {139} (\bibinfo {year} {2018})},\ \Eprint
  {http://arxiv.org/abs/1801.08955} {arXiv:1801.08955 [gr-qc]} \BibitemShut
  {NoStop}%
%%CITATION = ARXIV:1801.08955;%%
\bibitem [{\citenamefont {Zhu}\ \emph {et~al.}(2014)\citenamefont {Zhu},
  \citenamefont {Zhang}, \citenamefont {Pellicer}, \citenamefont {Wang},\ and\
  \citenamefont {Abdalla}}]{Zhu:2014sya}%
  \BibitemOpen
  \bibfield  {author} {\bibinfo {author} {\bibfnamefont {Z.}~\bibnamefont
  {Zhu}}, \bibinfo {author} {\bibfnamefont {S.-J.}\ \bibnamefont {Zhang}},
  \bibinfo {author} {\bibfnamefont {C.~E.}\ \bibnamefont {Pellicer}}, \bibinfo
  {author} {\bibfnamefont {B.}~\bibnamefont {Wang}}, \ and\ \bibinfo {author}
  {\bibfnamefont {E.}~\bibnamefont {Abdalla}},\ }\href {\doibase
  10.1103/PhysRevD.90.044042, 10.1103/PhysRevD.90.049904} {\bibfield  {journal}
  {\bibinfo  {journal} {Phys. Rev.}\ }\textbf {\bibinfo {volume} {D90}},\
  \bibinfo {pages} {044042} (\bibinfo {year} {2014})},\ \bibinfo {note}
  {[Addendum: Phys. Rev.D90,no.4,049904(2014)]},\ \Eprint
  {http://arxiv.org/abs/1405.4931} {arXiv:1405.4931 [hep-th]} \BibitemShut
  {NoStop}%
%%CITATION = ARXIV:1405.4931;%%
\bibitem [{\citenamefont {Konoplya}\ and\ \citenamefont
  {Zhidenko}(2014{\natexlab{a}})}]{Konoplya:2014lha}%
  \BibitemOpen
  \bibfield  {author} {\bibinfo {author} {\bibfnamefont {R.~A.}\ \bibnamefont
  {Konoplya}}\ and\ \bibinfo {author} {\bibfnamefont {A.}~\bibnamefont
  {Zhidenko}},\ }\href {\doibase 10.1103/PhysRevD.90.064048} {\bibfield
  {journal} {\bibinfo  {journal} {Phys. Rev.}\ }\textbf {\bibinfo {volume}
  {D90}},\ \bibinfo {pages} {064048} (\bibinfo {year} {2014}{\natexlab{a}})},\
  \Eprint {http://arxiv.org/abs/1406.0019} {arXiv:1406.0019 [hep-th]}
  \BibitemShut {NoStop}%
%%CITATION = ARXIV:1406.0019;%%
\bibitem [{\citenamefont {Dias}\ \emph {et~al.}(2010)\citenamefont {Dias},
  \citenamefont {Figueras}, \citenamefont {Monteiro}, \citenamefont {Reall},\
  and\ \citenamefont {Santos}}]{Dias:2010eu}%
  \BibitemOpen
  \bibfield  {author} {\bibinfo {author} {\bibfnamefont {O.~J.~C.}\
  \bibnamefont {Dias}}, \bibinfo {author} {\bibfnamefont {P.}~\bibnamefont
  {Figueras}}, \bibinfo {author} {\bibfnamefont {R.}~\bibnamefont {Monteiro}},
  \bibinfo {author} {\bibfnamefont {H.~S.}\ \bibnamefont {Reall}}, \ and\
  \bibinfo {author} {\bibfnamefont {J.~E.}\ \bibnamefont {Santos}},\ }\href
  {\doibase 10.1007/JHEP05(2010)076} {\bibfield  {journal} {\bibinfo  {journal}
  {JHEP}\ }\textbf {\bibinfo {volume} {05}},\ \bibinfo {pages} {076} (\bibinfo
  {year} {2010})},\ \Eprint {http://arxiv.org/abs/1001.4527} {arXiv:1001.4527
  [hep-th]} \BibitemShut {NoStop}%
%%CITATION = ARXIV:1001.4527;%%
\bibitem [{\citenamefont {Cardoso}\ \emph
  {et~al.}(2009{\natexlab{a}})\citenamefont {Cardoso}, \citenamefont {Lemos},\
  and\ \citenamefont {Marques}}]{Cardoso:2010rz}%
  \BibitemOpen
  \bibfield  {author} {\bibinfo {author} {\bibfnamefont {V.}~\bibnamefont
  {Cardoso}}, \bibinfo {author} {\bibfnamefont {M.}~\bibnamefont {Lemos}}, \
  and\ \bibinfo {author} {\bibfnamefont {M.}~\bibnamefont {Marques}},\ }\href
  {\doibase 10.1103/PhysRevD.80.127502} {\bibfield  {journal} {\bibinfo
  {journal} {Phys. Rev.}\ }\textbf {\bibinfo {volume} {D80}},\ \bibinfo {pages}
  {127502} (\bibinfo {year} {2009}{\natexlab{a}})},\ \Eprint
  {http://arxiv.org/abs/1001.0019} {arXiv:1001.0019 [gr-qc]} \BibitemShut
  {NoStop}%
%%CITATION = ARXIV:1001.0019;%%
\bibitem [{\citenamefont {Konoplya}\ and\ \citenamefont
  {Zhidenko}(2014{\natexlab{b}})}]{Konoplya:2013sba}%
  \BibitemOpen
  \bibfield  {author} {\bibinfo {author} {\bibfnamefont {R.~A.}\ \bibnamefont
  {Konoplya}}\ and\ \bibinfo {author} {\bibfnamefont {A.}~\bibnamefont
  {Zhidenko}},\ }\href {\doibase 10.1103/PhysRevD.89.024011} {\bibfield
  {journal} {\bibinfo  {journal} {Phys. Rev.}\ }\textbf {\bibinfo {volume}
  {D89}},\ \bibinfo {pages} {024011} (\bibinfo {year} {2014}{\natexlab{b}})},\
  \Eprint {http://arxiv.org/abs/1309.7667} {arXiv:1309.7667 [hep-th]}
  \BibitemShut {NoStop}%
%%CITATION = ARXIV:1309.7667;%%
\bibitem [{\citenamefont {Herdeiro}\ \emph {et~al.}(2013)\citenamefont
  {Herdeiro}, \citenamefont {Degollado},\ and\ \citenamefont
  {Rúnarsson}}]{Herdeiro:2013pia}%
  \BibitemOpen
  \bibfield  {author} {\bibinfo {author} {\bibfnamefont {C.~A.~R.}\
  \bibnamefont {Herdeiro}}, \bibinfo {author} {\bibfnamefont {J.~C.}\
  \bibnamefont {Degollado}}, \ and\ \bibinfo {author} {\bibfnamefont {H.~F.}\
  \bibnamefont {Rúnarsson}},\ }\href {\doibase 10.1103/PhysRevD.88.063003}
  {\bibfield  {journal} {\bibinfo  {journal} {Phys. Rev.}\ }\textbf {\bibinfo
  {volume} {D88}},\ \bibinfo {pages} {063003} (\bibinfo {year} {2013})},\
  \Eprint {http://arxiv.org/abs/1305.5513} {arXiv:1305.5513 [gr-qc]}
  \BibitemShut {NoStop}%
%%CITATION = ARXIV:1305.5513;%%
\bibitem [{\citenamefont {Degollado}\ and\ \citenamefont
  {Herdeiro}(2014)}]{Degollado:2013bha}%
  \BibitemOpen
  \bibfield  {author} {\bibinfo {author} {\bibfnamefont {J.~C.}\ \bibnamefont
  {Degollado}}\ and\ \bibinfo {author} {\bibfnamefont {C.~A.~R.}\ \bibnamefont
  {Herdeiro}},\ }\href {\doibase 10.1103/PhysRevD.89.063005} {\bibfield
  {journal} {\bibinfo  {journal} {Phys. Rev.}\ }\textbf {\bibinfo {volume}
  {D89}},\ \bibinfo {pages} {063005} (\bibinfo {year} {2014})},\ \Eprint
  {http://arxiv.org/abs/1312.4579} {arXiv:1312.4579 [gr-qc]} \BibitemShut
  {NoStop}%
%%CITATION = ARXIV:1312.4579;%%
\bibitem [{\citenamefont {Sanchis-Gual}\ \emph {et~al.}(2016)\citenamefont
  {Sanchis-Gual}, \citenamefont {Degollado}, \citenamefont {Montero},
  \citenamefont {Font},\ and\ \citenamefont {Herdeiro}}]{Sanchis-Gual:2015lje}%
  \BibitemOpen
  \bibfield  {author} {\bibinfo {author} {\bibfnamefont {N.}~\bibnamefont
  {Sanchis-Gual}}, \bibinfo {author} {\bibfnamefont {J.~C.}\ \bibnamefont
  {Degollado}}, \bibinfo {author} {\bibfnamefont {P.~J.}\ \bibnamefont
  {Montero}}, \bibinfo {author} {\bibfnamefont {J.~A.}\ \bibnamefont {Font}}, \
  and\ \bibinfo {author} {\bibfnamefont {C.}~\bibnamefont {Herdeiro}},\ }\href
  {\doibase 10.1103/PhysRevLett.116.141101} {\bibfield  {journal} {\bibinfo
  {journal} {Phys. Rev. Lett.}\ }\textbf {\bibinfo {volume} {116}},\ \bibinfo
  {pages} {141101} (\bibinfo {year} {2016})},\ \Eprint
  {http://arxiv.org/abs/1512.05358} {arXiv:1512.05358 [gr-qc]} \BibitemShut
  {NoStop}%
%%CITATION = ARXIV:1512.05358;%%
\bibitem [{\citenamefont {Li}(2013)}]{Li:2012nd}%
  \BibitemOpen
  \bibfield  {author} {\bibinfo {author} {\bibfnamefont {R.}~\bibnamefont
  {Li}},\ }\href {\doibase 10.1140/epjc/s10052-012-2274-9} {\bibfield
  {journal} {\bibinfo  {journal} {Eur. Phys. J.}\ }\textbf {\bibinfo {volume}
  {C73}},\ \bibinfo {pages} {2274} (\bibinfo {year} {2013})},\ \Eprint
  {http://arxiv.org/abs/1207.1984} {arXiv:1207.1984 [hep-th]} \BibitemShut
  {NoStop}%
%%CITATION = ARXIV:1207.1984;%%
\bibitem [{\citenamefont {Li}(2012)}]{Li:2012rx}%
  \BibitemOpen
  \bibfield  {author} {\bibinfo {author} {\bibfnamefont {R.}~\bibnamefont
  {Li}},\ }\href {\doibase 10.1016/j.physletb.2012.07.015} {\bibfield
  {journal} {\bibinfo  {journal} {Phys. Lett.}\ }\textbf {\bibinfo {volume}
  {B714}},\ \bibinfo {pages} {337} (\bibinfo {year} {2012})},\ \Eprint
  {http://arxiv.org/abs/1205.3929} {arXiv:1205.3929 [gr-qc]} \BibitemShut
  {NoStop}%
%%CITATION = ARXIV:1205.3929;%%
\bibitem [{\citenamefont {Li}\ and\ \citenamefont {Zhao}(2015)}]{Li:2014fna}%
  \BibitemOpen
  \bibfield  {author} {\bibinfo {author} {\bibfnamefont {R.}~\bibnamefont
  {Li}}\ and\ \bibinfo {author} {\bibfnamefont {J.}~\bibnamefont {Zhao}},\
  }\href {\doibase 10.1016/j.physletb.2014.12.007} {\bibfield  {journal}
  {\bibinfo  {journal} {Phys. Lett.}\ }\textbf {\bibinfo {volume} {B740}},\
  \bibinfo {pages} {317} (\bibinfo {year} {2015})},\ \Eprint
  {http://arxiv.org/abs/1412.1527} {arXiv:1412.1527 [gr-qc]} \BibitemShut
  {NoStop}%
%%CITATION = ARXIV:1412.1527;%%
\bibitem [{\citenamefont {Li}\ \emph {et~al.}(2015{\natexlab{a}})\citenamefont
  {Li}, \citenamefont {Zhao},\ and\ \citenamefont {Zhang}}]{Li:2014gfg}%
  \BibitemOpen
  \bibfield  {author} {\bibinfo {author} {\bibfnamefont {R.}~\bibnamefont
  {Li}}, \bibinfo {author} {\bibfnamefont {J.-K.}\ \bibnamefont {Zhao}}, \ and\
  \bibinfo {author} {\bibfnamefont {Y.-M.}\ \bibnamefont {Zhang}},\ }\href
  {\doibase 10.1088/0253-6102/63/5/569} {\bibfield  {journal} {\bibinfo
  {journal} {Commun. Theor. Phys.}\ }\textbf {\bibinfo {volume} {63}},\
  \bibinfo {pages} {569} (\bibinfo {year} {2015}{\natexlab{a}})},\ \Eprint
  {http://arxiv.org/abs/1404.6309} {arXiv:1404.6309 [gr-qc]} \BibitemShut
  {NoStop}%
%%CITATION = ARXIV:1404.6309;%%
\bibitem [{\citenamefont {Li}\ and\ \citenamefont {Zhao}(2014)}]{Li:2014xxa}%
  \BibitemOpen
  \bibfield  {author} {\bibinfo {author} {\bibfnamefont {R.}~\bibnamefont
  {Li}}\ and\ \bibinfo {author} {\bibfnamefont {J.}~\bibnamefont {Zhao}},\
  }\href {\doibase 10.1140/epjc/s10052-014-3051-8} {\bibfield  {journal}
  {\bibinfo  {journal} {Eur. Phys. J.}\ }\textbf {\bibinfo {volume} {C74}},\
  \bibinfo {pages} {3051} (\bibinfo {year} {2014})},\ \Eprint
  {http://arxiv.org/abs/1403.7279} {arXiv:1403.7279 [gr-qc]} \BibitemShut
  {NoStop}%
%%CITATION = ARXIV:1403.7279;%%
\bibitem [{\citenamefont {Li}\ \emph {et~al.}(2015{\natexlab{b}})\citenamefont
  {Li}, \citenamefont {Tian}, \citenamefont {Zhang},\ and\ \citenamefont
  {Zhao}}]{Li:2015mqa}%
  \BibitemOpen
  \bibfield  {author} {\bibinfo {author} {\bibfnamefont {R.}~\bibnamefont
  {Li}}, \bibinfo {author} {\bibfnamefont {Y.}~\bibnamefont {Tian}}, \bibinfo
  {author} {\bibfnamefont {H.-b.}\ \bibnamefont {Zhang}}, \ and\ \bibinfo
  {author} {\bibfnamefont {J.}~\bibnamefont {Zhao}},\ }\href {\doibase
  10.1016/j.physletb.2015.09.073} {\bibfield  {journal} {\bibinfo  {journal}
  {Phys. Lett.}\ }\textbf {\bibinfo {volume} {B750}},\ \bibinfo {pages} {520}
  (\bibinfo {year} {2015}{\natexlab{b}})},\ \Eprint
  {http://arxiv.org/abs/1506.04267} {arXiv:1506.04267 [gr-qc]} \BibitemShut
  {NoStop}%
%%CITATION = ARXIV:1506.04267;%%
\bibitem [{\citenamefont {Tanabe}(2016)}]{Tanabe:2015isb}%
  \BibitemOpen
  \bibfield  {author} {\bibinfo {author} {\bibfnamefont {K.}~\bibnamefont
  {Tanabe}},\ }\href {\doibase 10.1088/0264-9381/33/12/125016} {\bibfield
  {journal} {\bibinfo  {journal} {Class. Quant. Grav.}\ }\textbf {\bibinfo
  {volume} {33}},\ \bibinfo {pages} {125016} (\bibinfo {year} {2016})},\
  \Eprint {http://arxiv.org/abs/1511.06059} {arXiv:1511.06059 [hep-th]}
  \BibitemShut {NoStop}%
%%CITATION = ARXIV:1511.06059;%%
\bibitem [{\citenamefont {Jansen}(2017)}]{Jansen:2017oag}%
  \BibitemOpen
  \bibfield  {author} {\bibinfo {author} {\bibfnamefont {A.}~\bibnamefont
  {Jansen}},\ }\href {\doibase 10.1140/epjp/i2017-11825-9} {\bibfield
  {journal} {\bibinfo  {journal} {Eur. Phys. J. Plus}\ }\textbf {\bibinfo
  {volume} {132}},\ \bibinfo {pages} {546} (\bibinfo {year} {2017})},\ \Eprint
  {http://arxiv.org/abs/1709.09178} {arXiv:1709.09178 [gr-qc]} \BibitemShut
  {NoStop}%
%%CITATION = ARXIV:1709.09178;%%
\bibitem [{\citenamefont {Cardoso}\ \emph
  {et~al.}(2018{\natexlab{a}})\citenamefont {Cardoso}, \citenamefont {Costa},
  \citenamefont {Destounis}, \citenamefont {Hintz},\ and\ \citenamefont
  {Jansen}}]{Cardoso:2017soq}%
  \BibitemOpen
  \bibfield  {author} {\bibinfo {author} {\bibfnamefont {V.}~\bibnamefont
  {Cardoso}}, \bibinfo {author} {\bibfnamefont {J.~L.}\ \bibnamefont {Costa}},
  \bibinfo {author} {\bibfnamefont {K.}~\bibnamefont {Destounis}}, \bibinfo
  {author} {\bibfnamefont {P.}~\bibnamefont {Hintz}}, \ and\ \bibinfo {author}
  {\bibfnamefont {A.}~\bibnamefont {Jansen}},\ }\href {\doibase
  10.1103/PhysRevLett.120.031103} {\bibfield  {journal} {\bibinfo  {journal}
  {Phys. Rev. Lett.}\ }\textbf {\bibinfo {volume} {120}},\ \bibinfo {pages}
  {031103} (\bibinfo {year} {2018}{\natexlab{a}})},\ \Eprint
  {http://arxiv.org/abs/1711.10502} {arXiv:1711.10502 [gr-qc]} \BibitemShut
  {NoStop}%
%%CITATION = ARXIV:1711.10502;%%
\bibitem [{\citenamefont {Cardoso}\ \emph
  {et~al.}(2018{\natexlab{b}})\citenamefont {Cardoso}, \citenamefont {Costa},
  \citenamefont {Destounis}, \citenamefont {Hintz},\ and\ \citenamefont
  {Jansen}}]{PhysRevD.98.104007}%
  \BibitemOpen
  \bibfield  {author} {\bibinfo {author} {\bibfnamefont {V.}~\bibnamefont
  {Cardoso}}, \bibinfo {author} {\bibfnamefont {J.~L.}\ \bibnamefont {Costa}},
  \bibinfo {author} {\bibfnamefont {K.}~\bibnamefont {Destounis}}, \bibinfo
  {author} {\bibfnamefont {P.}~\bibnamefont {Hintz}}, \ and\ \bibinfo {author}
  {\bibfnamefont {A.}~\bibnamefont {Jansen}},\ }\href {\doibase
  10.1103/PhysRevD.98.104007} {\bibfield  {journal} {\bibinfo  {journal} {Phys.
  Rev. D}\ }\textbf {\bibinfo {volume} {98}},\ \bibinfo {pages} {104007}
  (\bibinfo {year} {2018}{\natexlab{b}})}\BibitemShut {NoStop}%
\bibitem [{\citenamefont {Liu}\ \emph {et~al.}(2019)\citenamefont {Liu},
  \citenamefont {Tang}, \citenamefont {Destounis}, \citenamefont {Wang},
  \citenamefont {Papantonopoulos},\ and\ \citenamefont {Zhang}}]{Liu:2019lon}%
  \BibitemOpen
  \bibfield  {author} {\bibinfo {author} {\bibfnamefont {H.}~\bibnamefont
  {Liu}}, \bibinfo {author} {\bibfnamefont {Z.}~\bibnamefont {Tang}}, \bibinfo
  {author} {\bibfnamefont {K.}~\bibnamefont {Destounis}}, \bibinfo {author}
  {\bibfnamefont {B.}~\bibnamefont {Wang}}, \bibinfo {author} {\bibfnamefont
  {E.}~\bibnamefont {Papantonopoulos}}, \ and\ \bibinfo {author} {\bibfnamefont
  {H.}~\bibnamefont {Zhang}},\ }\href {\doibase 10.1007/JHEP03(2019)187}
  {\bibfield  {journal} {\bibinfo  {journal} {JHEP}\ }\textbf {\bibinfo
  {volume} {03}},\ \bibinfo {pages} {187} (\bibinfo {year} {2019})},\ \Eprint
  {http://arxiv.org/abs/1902.01865} {arXiv:1902.01865 [gr-qc]} \BibitemShut
  {NoStop}%
%%CITATION = ARXIV:1902.01865;%%
\bibitem [{\citenamefont {Destounis}(2019)}]{Destounis:2018qnb}%
  \BibitemOpen
  \bibfield  {author} {\bibinfo {author} {\bibfnamefont {K.}~\bibnamefont
  {Destounis}},\ }\href {\doibase 10.1016/j.physletb.2019.06.015} {\bibfield
  {journal} {\bibinfo  {journal} {Phys. Lett.}\ }\textbf {\bibinfo {volume}
  {B795}},\ \bibinfo {pages} {211} (\bibinfo {year} {2019})},\ \Eprint
  {http://arxiv.org/abs/1811.10629} {arXiv:1811.10629 [gr-qc]} \BibitemShut
  {NoStop}%
%%CITATION = ARXIV:1811.10629;%%
\bibitem [{\citenamefont {Iyer}\ and\ \citenamefont
  {Will}(1987)}]{Iyer:1986np}%
  \BibitemOpen
  \bibfield  {author} {\bibinfo {author} {\bibfnamefont {S.}~\bibnamefont
  {Iyer}}\ and\ \bibinfo {author} {\bibfnamefont {C.~M.}\ \bibnamefont
  {Will}},\ }\href {\doibase 10.1103/PhysRevD.35.3621} {\bibfield  {journal}
  {\bibinfo  {journal} {Phys. Rev.}\ }\textbf {\bibinfo {volume} {D35}},\
  \bibinfo {pages} {3621} (\bibinfo {year} {1987})}\BibitemShut {NoStop}%
%%CITATION = PHRVA,D35,3621;%%
\bibitem [{\citenamefont {{Lin}}\ and\ \citenamefont {{Qian}}(2016)}]{KaiLin1}%
  \BibitemOpen
  \bibfield  {author} {\bibinfo {author} {\bibfnamefont {K.}~\bibnamefont
  {{Lin}}}\ and\ \bibinfo {author} {\bibfnamefont {W.-L.}\ \bibnamefont
  {{Qian}}},\ }\href@noop {} {\bibfield  {journal} {\bibinfo  {journal} {ArXiv
  e-prints}\ } (\bibinfo {year} {2016})},\ \Eprint
  {http://arxiv.org/abs/1609.05948} {arXiv:1609.05948 [math.NA]} \BibitemShut
  {NoStop}%
\bibitem [{\citenamefont {Kodama}\ \emph {et~al.}(2009)\citenamefont {Kodama},
  \citenamefont {Konoplya},\ and\ \citenamefont {Zhidenko}}]{Kodama:2009rq}%
  \BibitemOpen
  \bibfield  {author} {\bibinfo {author} {\bibfnamefont {H.}~\bibnamefont
  {Kodama}}, \bibinfo {author} {\bibfnamefont {R.~A.}\ \bibnamefont
  {Konoplya}}, \ and\ \bibinfo {author} {\bibfnamefont {A.}~\bibnamefont
  {Zhidenko}},\ }\href {\doibase 10.1103/PhysRevD.79.044003} {\bibfield
  {journal} {\bibinfo  {journal} {Phys. Rev.}\ }\textbf {\bibinfo {volume}
  {D79}},\ \bibinfo {pages} {044003} (\bibinfo {year} {2009})},\ \Eprint
  {http://arxiv.org/abs/0812.0445} {arXiv:0812.0445 [hep-th]} \BibitemShut
  {NoStop}%
%%CITATION = ARXIV:0812.0445;%%
\bibitem [{\citenamefont {Wang}\ and\ \citenamefont
  {Herdeiro}(2014)}]{Wang:2014eha}%
  \BibitemOpen
  \bibfield  {author} {\bibinfo {author} {\bibfnamefont {M.}~\bibnamefont
  {Wang}}\ and\ \bibinfo {author} {\bibfnamefont {C.}~\bibnamefont
  {Herdeiro}},\ }\href {\doibase 10.1103/PhysRevD.89.084062} {\bibfield
  {journal} {\bibinfo  {journal} {Phys. Rev.}\ }\textbf {\bibinfo {volume}
  {D89}},\ \bibinfo {pages} {084062} (\bibinfo {year} {2014})},\ \Eprint
  {http://arxiv.org/abs/1403.5160} {arXiv:1403.5160 [gr-qc]} \BibitemShut
  {NoStop}%
%%CITATION = ARXIV:1403.5160;%%
\bibitem [{\citenamefont {Cardoso}\ \emph
  {et~al.}(2009{\natexlab{b}})\citenamefont {Cardoso}, \citenamefont {Miranda},
  \citenamefont {Berti}, \citenamefont {Witek},\ and\ \citenamefont
  {Zanchin}}]{Cardoso:2008bp}%
  \BibitemOpen
  \bibfield  {author} {\bibinfo {author} {\bibfnamefont {V.}~\bibnamefont
  {Cardoso}}, \bibinfo {author} {\bibfnamefont {A.~S.}\ \bibnamefont
  {Miranda}}, \bibinfo {author} {\bibfnamefont {E.}~\bibnamefont {Berti}},
  \bibinfo {author} {\bibfnamefont {H.}~\bibnamefont {Witek}}, \ and\ \bibinfo
  {author} {\bibfnamefont {V.~T.}\ \bibnamefont {Zanchin}},\ }\href {\doibase
  10.1103/PhysRevD.79.064016} {\bibfield  {journal} {\bibinfo  {journal} {Phys.
  Rev.}\ }\textbf {\bibinfo {volume} {D79}},\ \bibinfo {pages} {064016}
  (\bibinfo {year} {2009}{\natexlab{b}})},\ \Eprint
  {http://arxiv.org/abs/0812.1806} {arXiv:0812.1806 [hep-th]} \BibitemShut
  {NoStop}%
%%CITATION = ARXIV:0812.1806;%%
\bibitem [{\citenamefont {Konoplya}\ and\ \citenamefont
  {Stuchlík}(2017)}]{Konoplya:2017wot}%
  \BibitemOpen
  \bibfield  {author} {\bibinfo {author} {\bibfnamefont {R.~A.}\ \bibnamefont
  {Konoplya}}\ and\ \bibinfo {author} {\bibfnamefont {Z.}~\bibnamefont
  {Stuchlík}},\ }\href {\doibase 10.1016/j.physletb.2017.06.015} {\bibfield
  {journal} {\bibinfo  {journal} {Phys. Lett.}\ }\textbf {\bibinfo {volume}
  {B771}},\ \bibinfo {pages} {597} (\bibinfo {year} {2017})},\ \Eprint
  {http://arxiv.org/abs/1705.05928} {arXiv:1705.05928 [gr-qc]} \BibitemShut
  {NoStop}%
%%CITATION = ARXIV:1705.05928;%%
\bibitem [{\citenamefont {Hod}(2012)}]{Hod:2013eea}%
  \BibitemOpen
  \bibfield  {author} {\bibinfo {author} {\bibfnamefont {S.}~\bibnamefont
  {Hod}},\ }\href {\doibase 10.1016/j.physletb.2012.06.043} {\bibfield
  {journal} {\bibinfo  {journal} {Phys. Lett.}\ }\textbf {\bibinfo {volume}
  {B713}},\ \bibinfo {pages} {505} (\bibinfo {year} {2012})},\ \Eprint
  {http://arxiv.org/abs/1304.6474} {arXiv:1304.6474 [gr-qc]} \BibitemShut
  {NoStop}%
%%CITATION = ARXIV:1304.6474;%%
\bibitem [{\citenamefont {Hod}(2017)}]{Hod:2017gvn}%
  \BibitemOpen
  \bibfield  {author} {\bibinfo {author} {\bibfnamefont {S.}~\bibnamefont
  {Hod}},\ }\href {\doibase 10.1140/epjc/s10052-017-4920-8} {\bibfield
  {journal} {\bibinfo  {journal} {Eur. Phys. J.}\ }\textbf {\bibinfo {volume}
  {C77}},\ \bibinfo {pages} {351} (\bibinfo {year} {2017})},\ \Eprint
  {http://arxiv.org/abs/1705.04726} {arXiv:1705.04726 [hep-th]} \BibitemShut
  {NoStop}%
%%CITATION = ARXIV:1705.04726;%%
\bibitem [{\citenamefont {Cardoso}\ and\ \citenamefont
  {Lemos}(2003)}]{Cardoso:2003sw}%
  \BibitemOpen
  \bibfield  {author} {\bibinfo {author} {\bibfnamefont {V.}~\bibnamefont
  {Cardoso}}\ and\ \bibinfo {author} {\bibfnamefont {J.~P.~S.}\ \bibnamefont
  {Lemos}},\ }\href {\doibase 10.1103/PhysRevD.67.084020} {\bibfield  {journal}
  {\bibinfo  {journal} {Phys. Rev.}\ }\textbf {\bibinfo {volume} {D67}},\
  \bibinfo {pages} {084020} (\bibinfo {year} {2003})},\ \Eprint
  {http://arxiv.org/abs/gr-qc/0301078} {arXiv:gr-qc/0301078 [gr-qc]}
  \BibitemShut {NoStop}%
%%CITATION = GR-QC/0301078;%%
\bibitem [{\citenamefont {Brill}\ and\ \citenamefont
  {Hayward}(1994)}]{Brill:1993tw}%
  \BibitemOpen
  \bibfield  {author} {\bibinfo {author} {\bibfnamefont {D.~R.}\ \bibnamefont
  {Brill}}\ and\ \bibinfo {author} {\bibfnamefont {S.~A.}\ \bibnamefont
  {Hayward}},\ }\href {\doibase 10.1088/0264-9381/11/2/008} {\bibfield
  {journal} {\bibinfo  {journal} {Class. Quant. Grav.}\ }\textbf {\bibinfo
  {volume} {11}},\ \bibinfo {pages} {359} (\bibinfo {year} {1994})},\ \Eprint
  {http://arxiv.org/abs/gr-qc/9304007} {arXiv:gr-qc/9304007 [gr-qc]}
  \BibitemShut {NoStop}%
%%CITATION = GR-QC/9304007;%%
\bibitem [{\citenamefont {Rendall}(2004)}]{Rendall:2003ks}%
  \BibitemOpen
  \bibfield  {author} {\bibinfo {author} {\bibfnamefont {A.~D.}\ \bibnamefont
  {Rendall}},\ }\href {\doibase 10.1007/s00023-004-0189-1} {\bibfield
  {journal} {\bibinfo  {journal} {Annales Henri Poincare}\ }\textbf {\bibinfo
  {volume} {5}},\ \bibinfo {pages} {1041} (\bibinfo {year} {2004})},\ \Eprint
  {http://arxiv.org/abs/gr-qc/0312020} {arXiv:gr-qc/0312020 [gr-qc]}
  \BibitemShut {NoStop}%
%%CITATION = GR-QC/0312020;%%
\bibitem [{\citenamefont {Lopez-Ortega}(2006)}]{LopezOrtega:2006my}%
  \BibitemOpen
  \bibfield  {author} {\bibinfo {author} {\bibfnamefont {A.}~\bibnamefont
  {Lopez-Ortega}},\ }\href {\doibase 10.1007/s10714-006-0335-9} {\bibfield
  {journal} {\bibinfo  {journal} {Gen. Rel. Grav.}\ }\textbf {\bibinfo {volume}
  {38}},\ \bibinfo {pages} {1565} (\bibinfo {year} {2006})},\ \Eprint
  {http://arxiv.org/abs/gr-qc/0605027} {arXiv:gr-qc/0605027 [gr-qc]}
  \BibitemShut {NoStop}%
%%CITATION = GR-QC/0605027;%%
\bibitem [{\citenamefont {Vasy}(2010)}]{VasydS}%
  \BibitemOpen
  \bibfield  {author} {\bibinfo {author} {\bibfnamefont {A.}~\bibnamefont
  {Vasy}},\ }\href@noop {} {\bibfield  {journal} {\bibinfo  {journal} {Advances
  in Mathematics}\ }\textbf {\bibinfo {volume} {223}},\ \bibinfo {pages} {49}
  (\bibinfo {year} {2010})}\BibitemShut {NoStop}%
\bibitem [{\citenamefont {Brady}\ \emph {et~al.}(1997)\citenamefont {Brady},
  \citenamefont {Chambers}, \citenamefont {Krivan},\ and\ \citenamefont
  {Laguna}}]{Brady:1996za}%
  \BibitemOpen
  \bibfield  {author} {\bibinfo {author} {\bibfnamefont {P.~R.}\ \bibnamefont
  {Brady}}, \bibinfo {author} {\bibfnamefont {C.~M.}\ \bibnamefont {Chambers}},
  \bibinfo {author} {\bibfnamefont {W.}~\bibnamefont {Krivan}}, \ and\ \bibinfo
  {author} {\bibfnamefont {P.}~\bibnamefont {Laguna}},\ }\href {\doibase
  10.1103/PhysRevD.55.7538} {\bibfield  {journal} {\bibinfo  {journal} {Phys.
  Rev.}\ }\textbf {\bibinfo {volume} {D55}},\ \bibinfo {pages} {7538} (\bibinfo
  {year} {1997})},\ \Eprint {http://arxiv.org/abs/gr-qc/9611056}
  {arXiv:gr-qc/9611056 [gr-qc]} \BibitemShut {NoStop}%
%%CITATION = GR-QC/9611056;%%
\bibitem [{\citenamefont {Brady}\ \emph {et~al.}(1999)\citenamefont {Brady},
  \citenamefont {Chambers}, \citenamefont {Laarakkers},\ and\ \citenamefont
  {Poisson}}]{Brady:1999wd}%
  \BibitemOpen
  \bibfield  {author} {\bibinfo {author} {\bibfnamefont {P.~R.}\ \bibnamefont
  {Brady}}, \bibinfo {author} {\bibfnamefont {C.~M.}\ \bibnamefont {Chambers}},
  \bibinfo {author} {\bibfnamefont {W.~G.}\ \bibnamefont {Laarakkers}}, \ and\
  \bibinfo {author} {\bibfnamefont {E.}~\bibnamefont {Poisson}},\ }\href
  {\doibase 10.1103/PhysRevD.60.064003} {\bibfield  {journal} {\bibinfo
  {journal} {Phys. Rev.}\ }\textbf {\bibinfo {volume} {D60}},\ \bibinfo {pages}
  {064003} (\bibinfo {year} {1999})},\ \Eprint
  {http://arxiv.org/abs/gr-qc/9902010} {arXiv:gr-qc/9902010 [gr-qc]}
  \BibitemShut {NoStop}%
%%CITATION = GR-QC/9902010;%%
\bibitem [{\citenamefont {Hod}(2018)}]{Hod:2018fet}%
  \BibitemOpen
  \bibfield  {author} {\bibinfo {author} {\bibfnamefont {S.}~\bibnamefont
  {Hod}},\ }\href {\doibase 10.1016/j.physletb.2018.09.039} {\bibfield
  {journal} {\bibinfo  {journal} {Phys. Lett.}\ }\textbf {\bibinfo {volume}
  {B786}},\ \bibinfo {pages} {217} (\bibinfo {year} {2018})},\ \Eprint
  {http://arxiv.org/abs/1808.04077} {arXiv:1808.04077 [gr-qc]} \BibitemShut
  {NoStop}%
%%CITATION = ARXIV:1808.04077;%%
\bibitem [{\citenamefont {Luna}\ \emph {et~al.}(2019)\citenamefont {Luna},
  \citenamefont {Zilhão}, \citenamefont {Cardoso}, \citenamefont {Costa},\
  and\ \citenamefont {Natário}}]{Luna:2018jfk}%
  \BibitemOpen
  \bibfield  {author} {\bibinfo {author} {\bibfnamefont {R.}~\bibnamefont
  {Luna}}, \bibinfo {author} {\bibfnamefont {M.}~\bibnamefont {Zilhão}},
  \bibinfo {author} {\bibfnamefont {V.}~\bibnamefont {Cardoso}}, \bibinfo
  {author} {\bibfnamefont {J.~L.}\ \bibnamefont {Costa}}, \ and\ \bibinfo
  {author} {\bibfnamefont {J.}~\bibnamefont {Natário}},\ }\href {\doibase
  10.1103/PhysRevD.99.064014} {\bibfield  {journal} {\bibinfo  {journal} {Phys.
  Rev.}\ }\textbf {\bibinfo {volume} {D99}},\ \bibinfo {pages} {064014}
  (\bibinfo {year} {2019})},\ \Eprint {http://arxiv.org/abs/1810.00886}
  {arXiv:1810.00886 [gr-qc]} \BibitemShut {NoStop}%
%%CITATION = ARXIV:1810.00886;%%
\end{thebibliography}%

\end{document}